\documentclass[preprint]{JHEP3} 
\usepackage{epsfig,multicol,bbm}
\usepackage{amsmath,epsfig,amssymb,bm}
\newcommand\fverb{\setbox\fverbbox=\hbox\bgroup\verb}
\newcommand\fverbdo{\egroup\medskip\noindent%
			\fbox{\unhbox\fverbbox}\ }
\newcommand\fverbit{\egroup\item[\fbox{\unhbox\fverbbox}]}
\newbox\fverbbox

\title{\mbox{}\\[10pt] Relativistic corrections to
the axial vector and vector currents in the
$\bm{\overline{b}c}$ meson system
at order $\bm{\alpha_s}$
}

\author{Jungil Lee and WenLong Sang \\
	Department of Physics, Korea University, Seoul 136-701, Korea\\
\email{jungil@korea.ac.kr},\email{swlong@korea.ac.kr}}
\author{Seyong~Kim \\
Department of Physics, Sejong University, Seoul 143-747, Korea\\
\email{skim@sejong.ac.kr}}

\received{\today} 		
\accepted{\today}		
\abstract{
We compute the short distance coefficients for the NRQCD factorization
formulas of the meson-to-vacuum matrix elements for the axial vector and
vector parts of the charged weak current in the $S$-wave spin-singlet and
-triplet $\bar{b}c$ mesons, respectively. The computation is carried out
to order $\alpha_s$ including relativistic corrections of all orders in
$\bm{q}^{2n}$, where $\bm{q}$ is the relative momentum of the $\bar{b}$
and $c$ in the meson rest frame. The relativistic corrections at
order $\alpha_s$ are new. The results reveal that the relativistic
corrections to the leptonic decay rate of the $B_c$ meson at order
$\alpha_s$ or less converge rapidly, which shows a strong contrast
to the uncomfortably large corrections of order $\alpha_s^2|\bm{q}|^0$.
The short distance coefficients listed in this paper can be employed to
compute the resummation of relativistic corrections to the phenomenological
measurables that involve $B_c$ and $B_c^*$ production and decay.}
\begin{document}
\section{Introduction\label{sec:introduction}}
Among various quarkonia including charmonia and bottomonia, the
bound state of a $\bar{b}c$ pair is a distinct heavy quarkonium system
composed of two different heavy quark flavors.
As the spin-singlet $S$-wave bound state of the $\bar{b}c$ pair,
the $B_c$ meson was discovered by the CDF Collaboration at the
Fermilab Tevatron through the decay mode $B_c\to
J/\psi+\ell^++\nu_\ell$ \cite{Abe:1998wi} and the $S$-wave
spin-triplet state $B_c^*$ has not been observed, yet\footnote{$B_c^+$
  is the bound state of a $\bar{b}c$ pair and $B_c^-$ is that of
  $\bar{c}b$. Throughout this paper, $B_c$ denotes $B_c^+$. However,
  our analysis can be equally applied to the charge-conjugate state $B_c^-$.}.
The $\bar{b}c$ meson offers a unique laboratory for the nonrelativistic quantum
chromodynamics (NRQCD) factorization framework \cite{Bodwin:1994jh} because
the typical heavy quark ($Q$) velocity $v_Q$ in the $\bar{b}c$ meson lies
between that ($v_c^2\sim 0.3$) in the charmonium system and that ($v_b^2\sim 0.1$)
in the bottomonium system. In addition, unlike the spin-triplet $S$-wave
$Q\bar{Q}$ mesons, $J/\psi$ and $\Upsilon$ that can decay into lepton pairs
through the electromagnetic current, the decay of the $B_c$ meson can proceed
only through the charged weak current and various dynamics play their roles
in the decay of the $\bar{b}c$ meson \cite{Isgur:1988gb,Lusignoli:1990ky,%
Chang:1992pt,Scora:1995ty,Beneke:1996xe,Kiselev:2000pp}.
As of now, except for the mass, $m_{B_c}=6.277\pm 0.006$\, GeV, and the life time,
$\tau_{B_c}=(0.45\pm 0.04)\times 10^{-12}$\,s, for the spin-singlet $S$-wave state
$B_c$, little is known about various properties of the $\bar{b}c$ bound states
experimentally \cite{Nakamura:2010zzi}. In near future, one may probe many
unknown properties of the bound states in detail as the CERN Large Hadron
Collider accumulates orders of magnitude larger number of events
than the currently available data \cite{Brambilla:2004wf} for the
$\bar{b}c$ bound states. Therefore, it is desirable and necessary to
achieve better accuracies in theoretical predictions.

Earlier theoretical studies on the $\bar{b}c$ bound states cover the
spectroscopy \cite{Eichten:1980mw,Godfrey:1985xj,Kwong:1990am,Eichten:1994gt,%
Kiselev:1994rc} and the production mechanism at colliders \cite{Chang:1991bp,%
Chang:1992bb,Chang:1992jb,Cheung:1993qi,Braaten:1993jn}.
In order to achieve better accuracies in the predictions for the $\bar{b}c$-meson
production and decay, it is necessary to know accurate values for
the decay constants. The NRQCD factorization formula is useful in making
a systematic series expansion of the decay constant for the $\bar{b}c$ meson
system in powers of $v_Q$. The NRQCD factorization theorems have been proved
for the electromagnetic and light hadronic decays of heavy
quarkonia \cite{Bodwin:1994jh} and for a few exclusive production processes
of heavy quarkonia \cite{Bodwin:2008nf,Bodwin:2009cb,Bodwin:2010fi}.
The NRQCD factorization formulas for the hadronic part of the
charged weak current that involves the decays of $B_c$ and $B_c^*$ are
similar to those for the electromagnetic decays of $Q\bar{Q}$ mesons.
Braaten and Fleming computed the one-loop
QCD corrections to the $B_c$ decay constant in the static limit $v_Q=0$
and the relativistic corrections of relative order
$\alpha_s^0v_Q^2$ \cite{Braaten:1995ej} by computing the short distance
coefficients of the NRQCD factorization formula for the decay constant.
Based on the same strategy, Hwang and Kim calculated the
$B_c^*$ counterparts \cite{Hwang:1999fc}. Two-loop QCD correction
to the short distance coefficient for axial vector current involving
$B_c$ decay constant was calculated by Onishchenko and
Veretin \cite{Onishchenko:2003ui}, which shows uncomfortably large
correction  like those for the leptonic decays of the spin-triplet $S$-wave
quarkonia $J/\psi$ and $\Upsilon$ \cite{Beneke:1997jm,Czarnecki:1997vz}.
In fact, according to the velocity-scaling rules of NRQCD \cite{Bodwin:1994jh},
the corrections of relative orders $v_Q^4$ and $\alpha_s v_Q^2$ should be
equally important as that of relative order $\alpha_s^2v_Q^0$.
In addition, the large separation between $m_b$ and $m_c$, where $m_Q$
is the mass of the heavy quark $Q=b$ or $c$, in the $\bar{b}c$ system
gives rise to factors of $\log (m_b/m_c)$ in the one-loop corrections
\cite{Braaten:1995ej,Hwang:1999fc,Onishchenko:2003ui}, which
may potentially deteriorate the convergence of the power expansion in $v_Q$.
Therefore, it is worthwhile to check if such large corrections indeed
arise when one includes the relativistic effects.

In this paper, we compute the relativistic corrections to the NRQCD
factorization formulas of the meson-to-vacuum matrix elements for
the axial vector and vector parts of the charged weak current in
the $S$-wave spin-singlet and -triplet $\bar{b}c$ mesons, respectively.
The calculation is carried out at first order of the strong coupling
$\alpha_s$ including relativistic corrections to all orders in $v_Q$.
The short distance coefficients for the NRQCD factorization formula
are usually obtained after subtracting the infrared (IR)-sensitive contributions
of the NRQCD correction from the full QCD corrections by perturbative matching.
It requires laborious bookkeeping of the Feynman rules in
the NRQCD perturbation theory which grows tremendously extensive as the order
in $v_Q^{2n}$ increases \cite{Luke:1997ys,Griesshammer:1997wz}.
Instead, we use a new method
introduced recently in \cite{Bodwin:2008vp} to compute the relativistic
corrections to the short distance coefficients at order $\alpha_s$ covering
all orders in $v_Q^{2n}$. The method integrates out the temporal component
of the loop momentum by contour
integration in the one-loop corrections to the full QCD amplitude and, then,
expands the integrands in powers of the external momenta divided by
$m_Q$ and the spatial components of the loop momentum divided by $m_Q$.
In comparison with, so called, the {\it method of region} given in
\cite{Beneke:1997zp}, this new method is particularly useful in
computing relativistic corrections of higher orders in $v_Q$
and even makes it possible to find the closed form of the expression
that includes the relativistic corrections resummed to all
orders in $v_Q^{2n}$ at one loop.

This paper is organized as follows.
In section~\ref{sec:matching}, we discuss the perturbative matching of NRQCD
onto QCD at one loop. Kinematics of the problem and the definitions of the
variables that are useful in computing the short distance coefficients
are given in section~\ref{sec:kinematics}. Section~\ref{sec:shortdistance}
contains the strategy and detailed formulas to compute the short distance
coefficients. We compute the QCD one-loop corrections in section~\ref{sec:QCD}
followed by the NRQCD corrections in section~\ref{sec:NRQCD}. Our final results
for the short distance coefficients are listed in section~\ref{sec:results}
and we summarize in section~\ref{sec:summary}.
In appendices, we provide the formulas for the
tensor-integral reduction and list the values for the loop integrals
that appear in the one-loop correction to the QCD and NRQCD amplitudes.
\section{Perturbative matching to all orders in $\bm{v}$\label{sec:matching}}
We define the hadronic parts of the weak decay amplitudes
$i {\cal A}^\mu_{5,{B_c}}$ and $i {\cal A}^\mu_{{B_c^*}}$
for $B_c$ and $B_c^*$ as the meson-to-vacuum matrix elements of
the axial vector current ${\overline{b}}{\gamma}^{\mu}{\gamma}_5 c$ and
the vector current ${\overline{b}}{\gamma}^{\mu}c$, respectively as\footnote{
See, for example, \cite{Lusignoli:1990ky}.}:
\begin{subequations}
\label{a1ab}
\begin{eqnarray}
i {\cal A}^\mu_{5,{B_c}}&\equiv&
\langle 0|\, {\overline{b}}{\gamma}^{\mu}{\gamma}_5 c\,
  |B_c\rangle = i f_{B_c}K^{\mu},
\label{a1a}\\
i {\cal A}^\mu_{{B_c^*}}&\equiv&
\langle 0|\, {\overline{b}}{\gamma}^{\mu}c \,
|B_c^*\rangle = i
f_{B^*_c}m_{B^*_c}{\varepsilon}^{\mu},
\label{a1b}
\end{eqnarray}
\end{subequations}
where $b$ and $c$ are the Dirac field operators for the bottom quark
and charm quark, respectively. The amplitudes (\ref{a1ab}) are scaled by the
leptonic decay constants $f_{H}$ for $H=B_c$ and $B_c^*$.
Here, $K$ is the meson momentum,
$\varepsilon$ is the polarization vector of the $B_c^*$ meson,
$m_{B_c}=$ 6.277\,(6)\,GeV \cite{Nakamura:2010zzi} and
$m_{B_c^*}=$ 6.330\,(7)(2)(6)\,GeV \cite{Gregory:2009hq}
are the masses of the $B_c$ and $B_c^*$ mesons.
The quarkonium state $|H\rangle$ in
(\ref{a1ab}) for $H=B_c$ and $B_c^*$ is normalized relativistically:
$\langle H(K')|H(K)\rangle=2\,K^0(2\pi)^3\delta^{(3)}(\bm{K}'-\bm{K})$.
In the meson rest frame, the matrix elements in (\ref{a1ab})
become simple: Because $K=(m_H,\bm{0})$ in this frame,
only the $0$-th component survives in the matrix element (\ref{a1a})
for the $B_c$. Due to the transverse condition
$\varepsilon\cdot K=0$ for the $B_c^*$, $\varepsilon=(0,\bm{\varepsilon})$
in this frame and, therefore, only the spatial components are nonvanishing
in the matrix element (\ref{a1b}) for the $B_c^*$.

According to the NRQCD factorization \cite{Bodwin:1994jh},
we can write the nonvanishing components of $i {\cal A}^\mu_{5,{B_c}}$
and $i {\cal A}^\mu_{{B_c^*}}$ in the rest frame
of the $\bar{b}c$ bound states as
\begin{subequations}
\label{me-org}
\begin{eqnarray}
i {\cal A}^0_{5,{B_c}}&=&
\sqrt{2m_{\!B_c}}
\sum_n
P_n \langle 0|\mathcal{O}_n|B_c \rangle,
\\
i {\cal A}^i_{{B_c^*}}&=&
\sqrt{2m_{\!B_c^*}}
\sum_n
V_n \langle 0|\mathcal{O}^i_n|B_c^* \rangle,
\end{eqnarray}
\end{subequations}
where $P_n$ and $V_n$ are the short distance coefficients
and $\mathcal{O}_n$ and $\mathcal{O}_n^i$ are the NRQCD operators.
The operator matrix elements in (\ref{me-org}) are regularized
dimensionally in $d=4-2\epsilon$ space-time dimensions.
The overall factor $\sqrt{2m_H}$ for $H=B_c$ and $B_c^*$ in
(\ref{me-org}) has been taken out because the state
$|H\rangle$ in (\ref{me-org}) is normalized nonrelativistically:
$\langle H(\bm{K}')|H(\bm{K})\rangle=(2\pi)^3\delta^{(3)}(\bm{K}'-\bm{K})$,
while the amplitudes on the left sides of (\ref{me-org}) have
the relativistic normalization for the quarkonium $H$ like those in
(\ref{a1ab}).

The main purpose of this paper is to compute the short distance coefficients
$P_n$ and $V_n$ in (\ref{me-org}) that correspond to $\bar{b}c$
color-singlet operators at order $\alpha_s$. These coefficients can be
determined by the matching equations
\begin{subequations}
\label{matching}
\begin{eqnarray}
i {\cal A}^0_{5,\bar{b}c_1}&=&
\sum_n
P_n \langle 0|\mathcal{O}_n|\bar{b}c_1 \rangle,
\\
i {\cal A}^i_{\bar{b}c_1}&=&
\sum_n
V_n \langle 0|\mathcal{O}^i_n|\bar{b}c_1 \rangle,
\end{eqnarray}
\end{subequations}
which is the statement of NRQCD factorization for the perturbative
color-singlet $\bar{b}c$ state.
Here, $\bar{b}c_1$ denotes the color-singlet $\bar{b}c$ pair
whose invariant mass is the same as the meson mass. Throughout
this paper, we suppress the factor $\sqrt{N_c}$ that comes from
the implicit color trace in $i {\cal A}^0_{5,\bar{b}c_1}$
and $i {\cal A}^i_{\bar{b}c_1}$, where $N_c=3$ is the number of
colors. Note that the coefficients $P_n$ and $V_n$
in (\ref{matching}) are identical to those in (\ref{me-org})
because the short distance coefficients must not depend on the
long distance nature of the heavy quarkonium state.
While the amplitudes (\ref{me-org}) contain nonperturbative
quantities, the amplitudes (\ref{matching}) are calculable perturbatively.
However, it is possible that the amplitudes (\ref{matching}) acquire
singularities in the IR or ultraviolet (UV) regions at order $\alpha_s$
or higher. These divergences are to be regularized dimensionally.
The matching equations that contain the terms upto order $\alpha_s$ are
\begin{subequations}
\label{matching-1}
\begin{eqnarray}
i {\cal A}^{0(0)}_{5,\bar{b}c_1}
+i {\cal A}^{0(1)}_{5,\bar{b}c_1}
&=&
\sum_n
(P_n^{(0)}+P_n^{(1)}) \langle 0|\mathcal{O}_n|\bar{b}c_1 \rangle^{(0)}
+
\sum_n
P_n^{(0)} \langle 0|\mathcal{O}_n|\bar{b}c_1 \rangle^{(1)},
\\
i {\cal A}^{i(0)}_{\bar{b}c_1}
+i {\cal A}^{i(1)}_{\bar{b}c_1}
&=&
\sum_n
(V_n^{(0)}+V_n^{(1)}) \langle 0|\mathcal{O}^i_n|\bar{b}c_1 \rangle^{(0)}
+
\sum_n
V_n^{(0)} \langle 0|\mathcal{O}^i_n|\bar{b}c_1 \rangle^{(1)},
\end{eqnarray}
\end{subequations}
where the superscripts $(0)$ and $(1)$ indicate the order in $\alpha_s$.
In the first sum of each line in (\ref{matching-1}), only
color-singlet $\bar{b}c$ operators contribute, while in the next sum
additional operators may enter once they mix with color-singlet
$\bar{b}c$ operators under one-loop QCD corrections.

Through order $\alpha_s$, the NRQCD amplitudes are defined by
\begin{subequations}
\label{matching-2}
\begin{eqnarray}
{[}i {\cal A}^{0(j)}_{5,\bar{b}c_1}]_{\rm NRQCD}
&=&
\sum_n
P_n^{(0)}\langle 0|\mathcal{O}_n|\bar{b}c_1 \rangle^{(j)},
\\
{[}i {\cal A}^{i(j)}_{\bar{b}c_1}]_{\rm NRQCD}
&=&
\sum_n
V_n^{(0)}\langle 0|\mathcal{O}^i_n|\bar{b}c_1 \rangle^{(j)},
\end{eqnarray}
\end{subequations}
where $j=0$ or 1. At order $\alpha_s^0$, the NRQCD matrix elements
$\langle 0|\mathcal{O}_n|\bar{b}c_1 \rangle^{(0)}$
and
$\langle 0|\mathcal{O}^i_n|\bar{b}c_1 \rangle^{(0)}$
are finite and the short distance coefficients
$P_n^{(0)}$ and $V_n^{(0)}$ can be determined from the identities
\begin{subequations}
\label{matching-3}
\begin{eqnarray}
\label{matching-3-5}
{[}i {\cal A}^{0(0)}_{5,\bar{b}c_1}]_{\rm NRQCD}
&=&
i {\cal A}^{0(0)}_{5,\bar{b}c_1}
=\sum_n
P_n^{(0)}\langle 0|\mathcal{O}_n|\bar{b}c_1 \rangle^{(0)},
\\
\label{matching-3-1}
{[}i {\cal A}^{i(0)}_{\bar{b}c_1}]_{\rm NRQCD}
&=&
i {\cal A}^{i(0)}_{\bar{b}c_1}
=\sum_n
V_n^{(0)}\langle 0|\mathcal{O}^i_n|\bar{b}c_1 \rangle^{(0)}.
\end{eqnarray}
\end{subequations}
Each expansion of (\ref{matching-3}) is a power series
in $\bm{q}/m_Q$, where $\bm{q}$ is half the relative three-momentum of
the $\bar{b}$ and $c$ in the center-of-momentum (CM) frame of the
$\bar{b}c$ pair. However, at order $\alpha_s$, the short distance
coefficients $P_n^{(1)}$ and $V_n^{(1)}$ must be determined after
subtracting the long distance contributions that are contained in
the order-$\alpha_s$ matrix elements
$\langle 0|\mathcal{O}_n|\bar{b}c_1 \rangle^{(1)}$ and
$\langle 0|\mathcal{O}^i_n|\bar{b}c_1 \rangle^{(1)}$, which include the
contributions from the potential, soft, and ultrasoft regions:
\begin{subequations}
\label{matching-4}
\begin{eqnarray}
\label{matching-4-5}
i {\cal A}^{0(1)}_{5,\bar{b}c_1}
-
{[}i {\cal A}^{0(1)}_{5,\bar{b}c_1}]_{\rm NRQCD}
&=&
\sum_n
P_n^{(1)}\langle 0|\mathcal{O}_n|\bar{b}c_1 \rangle^{(0)},
\\
\label{matching-4-1}
i {\cal A}^{i(1)}_{\bar{b}c_1}
-
{[}i {\cal A}^{i(1)}_{\bar{b}c_1}]_{\rm NRQCD}
&=&
\sum_n
V_n^{(1)}\langle 0|\mathcal{O}^i_n|\bar{b}c_1 \rangle^{(0)}.
\end{eqnarray}
\end{subequations}
In this way, one can determine the order-$\alpha_s$ short distance
coefficients $P_n^{(1)}$ and $V_n^{(1)}$, which are free of
IR sensitivity.

The computation of
${[}i {\cal A}^{0(1)}_{5,\bar{b}c_1}]_{\rm NRQCD}$
and
${[}i {\cal A}^{i(1)}_{\bar{b}c_1}]_{\rm NRQCD}$ is very complicated
because it involves operators and interactions of all orders in $v_Q$.
Fortunately, the authors of \cite{Bodwin:2008vp} recently introduced
a way to compute the NRQCD amplitudes
$[i {\cal A}^{0(1)}_{5,\bar{b}c_1}]_{\rm NRQCD}$
and
$[i {\cal A}^{i(1)}_{\bar{b}c_1}]_{\rm NRQCD}$
directly from the full QCD counterparts
$i {\cal A}^{0(1)}_{5,\bar{b}c_1}$
and
$i {\cal A}^{i(1)}_{\bar{b}c_1}$ by expanding the integrands in powers of
the momentum divided by $m_Q$. Before the expansion, temporal component
of the loop momentum has been integrated out, using contour integration,
in order to avoid the generation of ill defined pinch singularities that
may develop if one expands the heavy quark propagators too early.
The integrands for the integration over the remaining spatial
components of the loop momentum are then expanded in powers of the external
momenta divided by $m_Q$ and the spatial components of the loop momenta
divided by $m_Q$. Divergent integrals over the spatial components of the
loop momenta are regularized dimensionally while scaleless
power-divergent integrals are dropped in accordance with the dimensional
regularization scheme. As the last step, remaining UV
divergences are renormalized according to the modified minimal
subtraction ($\overline{\textrm{MS}}$) scheme. In this work, we apply
this method to compute the
NRQCD amplitudes ${[}i {\cal A}^{0(1)}_{5,\bar{b}c_1}]_{\rm NRQCD}$
and ${[}i {\cal A}^{i(1)}_{\bar{b}c_1}]_{\rm NRQCD}$
and determine the short distance coefficients
$P_n^{(1)}$ and $V_n^{(1)}$ from (\ref{matching-4}).
\section{Kinematics and notations\label{sec:kinematics}}
In this section, we define notations for the kinematics of the problem.
We take $p_{c}$ and $p_{\bar{b}}$ to be the momenta of the incoming heavy
quark $c$ and heavy antiquark $\bar{b}$, respectively, which are on their
mass shells: $p_{c}^2=m_c^2$ and $p_{\bar{b}}^2=m_b^2$.
They are expressed as linear combinations of half the total
momentum $p=\tfrac{1}{2}(p_c+p_{\bar{b}})=\tfrac{1}{2}K$
and half their relative momentum $q=\tfrac{1}{2}(p_c-p_{\bar{b}})$:
\begin{subequations}
\begin{eqnarray}
p_c&=&p+q,
\\
p_{\bar{b}}&=&p-q.
\end{eqnarray}
\end{subequations}
In the CM frame of the $\bar{b}c$ pair, the momenta are given by
\begin{subequations}
\begin{eqnarray}
p_c&=&(E_c,\bm{q}),
\\
p_{\bar{b}}&=&(E_b,-\bm{q}),
\\
p&=&[\tfrac{1}{2}(E_c+E_b),\bm{0}],
\\
q&=&[\tfrac{1}{2}(E_c-E_b),\bm{q}],
\end{eqnarray}
\end{subequations}
where $E_c=(m_c^2+\bm{q}^2)^{1/2}$ and
$E_{b}=(m_b^2+\bm{q}^2)^{1/2}$. Note that
\begin{equation}
4 p \cdot q = (p_c+p_{\bar{b}}) \cdot (p_c-p_{\bar{b}})
=m_c^2-m_b^2\neq 0,
\end{equation}
unlike the case of the $Q\bar{Q}$ pair considered in
\cite{Bodwin:2008vp}.

For later use, it is convenient to define parameters $\delta$ which
is the magnitude of the three-momentum of $\bar{b}$ or $c$ and $e_c$
which is the energy of the charm quark scaled by the invariant mass
$\sqrt{4p^2}=E_c+E_b$ of the $\bar{b}c$ pair in the CM frame:
\begin{subequations}
\label{def-ab}
\begin{eqnarray}
\delta &=&
\frac{|\bm{q}|}{\sqrt{4p^2}}
=
 \frac{1}{2}
    \sqrt{\left[1 - \frac{(m_b+m_c)^2}{4 p^2}\right]
    \left[1 - \frac{(m_b -m_c)^2}{4 p^2}\right]},
\\
e_c &=&
\frac{E_c}{\sqrt{4p^2}}
=
 \frac{1}{2} \left(1 - \frac{m_b^2-m_c^2}{4p^2}\right).
\end{eqnarray}
\end{subequations}
Note that $\delta/e_c=|\bm{q}|/E_c$.
The following relations are also useful:
\begin{subequations}
\label{pbpc}
\begin{eqnarray}
p_{\bar{b}}\cdot p_c+m_bm_c
&=&(E_bE_c+m_bm_c)+\bm{q}^2
=2m_bm_c+{O}(\bm{q}^2),
\\
p_{\bar{b}}\cdot p_c-m_bm_c
&=&(E_bE_c-m_bm_c)+\bm{q}^2
=\frac{(m_b+m_c)^2}{2m_bm_c}\,\bm{q}^2+{O}(\bm{q}^4).
\end{eqnarray}
\end{subequations}

In the derivation of the full QCD amplitudes, we use
the Dirac spinors for the $c$ and $\bar{b}$ with the nonrelativistic
normalization. In the CM frame of the $\bar{b}c$ pair, they are
\begin{subequations}
\label{uv-spinors}
\begin{eqnarray}
 u_c (p_c) &=& \mathcal{N}_c
 \begin{pmatrix}
 (E_c+m_c)\xi_{c}
 \\
 \bm{q}\cdot\bm{\sigma}\xi_{c}
 \end{pmatrix},
\\
 v_c (p_{\bar{b}}) &=& \mathcal{N}_b
 \begin{pmatrix}
 -\bm{q}\cdot\bm{\sigma}\eta_{b}
\\
 (E_b+m_b)\eta_{b}
 \end{pmatrix},
\end{eqnarray}
\end{subequations}
where $\mathcal{N}_Q=[2E_Q(E_Q+m_Q)]^{-1/2}$ for
$Q=c$ and $b$, $\xi_{c}$ and  $\eta_{b}$ are Pauli spinors for the $c$
and $\bar{b}$, respectively. The spinors in (\ref{uv-spinors})
are convenient in making nonrelativistic expansions. The {\it threshold
expansion method} in \cite{Braaten:1996jt} and its dimensionally
regularized version in \cite{Braaten:1996rp} also use this form
except that the relativistic normalization is used.
To extract the spin-singlet and -triplet states from a full QCD amplitude
for the $\bar{b}c$ pair, one can also make use of the spin-projection
operators for those states. In \cite{Sang:2009zz,Bodwin:2010fi},
for example, one can find the spin-projection operators for the
spin-singlet and -triplet states of a heavy quark-antiquark pair with
different flavors.

We use the representation for
the Dirac matrices introduced in \cite{Braaten:1996rp}:
\begin{equation}
\label{Dirac-matrices}
\gamma^0=\begin{pmatrix}
\mathbbm{1}&\phantom{+}0\\
0&-\mathbbm{1}
\end{pmatrix},
\quad
\gamma^i=\begin{pmatrix}
\phantom{+}0&\phantom{+}\sigma^i\\
-\sigma^i&\phantom{+}0
\end{pmatrix},
\end{equation}
where $\mathbbm{1}$ is the identity matrix. In (\ref{Dirac-matrices})
$\gamma^i$ and the Pauli matrix $\sigma^i$ are defined for
$i=1,$ 2, $\cdots$, $d-1$.
The requirement of the Clifford algebra for the Dirac matrices in $d$
space-time dimensions
\begin{equation}
\label{anticommutation-relation}
\{\gamma^\mu,\gamma^\nu\}=2g^{\mu\nu}\mathbbm{1},
\end{equation}
for $\mu,\,\nu=0$, 1, 2, $\cdots$, $d-1$ forces the anticommutation
relations for the Pauli matrices,
\begin{equation}
\label{sigma-delta}
\{\sigma^i,\sigma^j\}=2\delta^{ij}\mathbbm{1},
\end{equation}
for $i,\, j=1$, 2, $\cdots$, $d-1$.
In our computation of the spin-singlet case, we encounter the loop
correction to the axial vector current. We carry out the Dirac algebra
by making use of the naive dimensional regularization, in which
$\gamma_5$ anticommutes with $\gamma^\mu$ for any indices $\mu$ in $d$
dimensions. This prescription is self consistent for the case
considered in this paper \cite{Larin:1993tq}. As commented in section
3 of \cite{Petrelli:1997ge}, the matrix representation of the Dirac
$\gamma_5$ that is consistent with the choice (\ref{Dirac-matrices})
is then
\begin{equation}
\label{gamma5}
\gamma_5=
\begin{pmatrix}
0&\phantom{+}\mathbbm{1}\\
\mathbbm{1}&\phantom{+}0
\end{pmatrix}.
\end{equation}
which guarantees $\{\gamma^\mu,\gamma_5\}=0$.
With the matrix representations for the spinors in (\ref{uv-spinors})
and with the set of Dirac matrices (\ref{Dirac-matrices}) and
(\ref{gamma5}), we can carry out the calculation consistent with naive
dimensional regularization.

For $d-1=3$, the Pauli matrices
satisfy the commutation relations
\begin{equation}
\label{commutation-relation}
[\sigma^i,\sigma^j]=2i\epsilon^{ijk}\sigma^k.
\end{equation}
However,
for the spatial dimensions greater than 3 the totally antisymmetric
combination of three Pauli matrices
$\{[\sigma^i,\sigma^j],\sigma^k\}$, which may arise in the threshold
expansion of the products of three or more Dirac matrices of
different spatial indices \cite{Braaten:1996jt},
is linearly independent of both $\mathbbm{1}$ and $\sigma^\ell$ for
$\ell=1$, 2, $\cdots$, $d-1$. And the reduction
\begin{equation}
\label{Pauli-reduction}
\{[\sigma^i,\sigma^j],\sigma^k\}=4i\epsilon^{ijk}\mathbbm{1}
\end{equation}
is allowed only at $3$ spatial dimensions~\cite{Braaten:1996rp}.
Therefore, unless divergent contributions disappear, we do not use
the reduction (\ref{Pauli-reduction}).
\section{Formulas for short distance coefficients\label{sec:shortdistance}}
Let us first classify the operators that appear in the
matching conditions (\ref{matching-3}) and (\ref{matching-4}).
For the spin-singlet $S$-wave case, only a single type of operators
$\mathcal{O}_{n}$ contributes and for the spin-triplet $S$-wave case,
there are two kinds of operators $\mathcal{O}^i_{An}$ and
$\mathcal{O}^i_{Bn}$:
\begin{subequations}
\label{operators}
\begin{eqnarray}
\mathcal{O}_{n}
&=&
\chi_b^\dagger (-\tfrac{i}{2}
\stackrel{\leftrightarrow}{\bm{\nabla}}
)^{2n}\psi_c,
\\
\mathcal{O}^i_{An}
&=&
\chi_b^\dagger (-\tfrac{i}{2}
\stackrel{\leftrightarrow}{\bm{\nabla}}
)^{2n}\sigma^i\psi_c,
\\
\mathcal{O}^i_{Bn}
&=&
\chi_b^\dagger (-\tfrac{i}{2}
\stackrel{\leftrightarrow}{\bm{\nabla}}
)^{2n-2}
(-\tfrac{i}{2}
\stackrel{\leftrightarrow}{{\nabla}}^i
)
(-\tfrac{i}{2}
\stackrel{\leftrightarrow}{\bm{\nabla}}
)\cdot\bm{\sigma}
\psi_c,
\end{eqnarray}
\end{subequations}
where $\psi_c$ is the Pauli spinor field that annihilates the
charm quark and $\chi_b^\dagger$ is the Pauli spinor field that
annihilates the antibottom quark.
The operators in (\ref{operators}) contain ordinary derivatives,
rather than covariant derivatives so that they are not gauge invariant.
We evaluate their matrix elements in the Coulomb gauge, in which
inclusion of the $\bar{b}c$ operators involving the gauge fields
brings in corrections of relative order $v_Q^4$~\cite{Bodwin:2010fi}.
While the operators $\mathcal{O}_n$ and $\mathcal{O}^i_{An}$ have
only the $S$-wave contributions, the operator $\mathcal{O}^i_{Bn}$ also
contains the $D$-wave contribution as well as the $S$-wave one.
The operator ${\cal O}_{Bn}^i$ can be decomposed into a
linear combination of ${\cal O}_{An}^i$ and the $D$-wave operator
${\cal O}_{Dn}^i$:
\begin{equation}
\label{OPerator-B}%
{\cal O}_{Bn}^i=
\frac{1}{d-1}
{\cal O}_{An}^i+
{\cal O}_{Dn}^i,
\end{equation}
where ${\cal O}_{Dn}^i$ is defined by
\begin{equation}
\label{OPerator-D}%
{\cal O}_{Dn}^i=
\chi_b^\dagger (-\tfrac{i}{2}\stackrel{\leftrightarrow}{\bm{\nabla}})^{2n-2}
\left[
(-\tfrac{i}{2}\stackrel{\leftrightarrow}{\nabla}^i)
(-\tfrac{i}{2}\stackrel{\leftrightarrow}{\bm{\nabla}})\cdot\bm{\sigma}
-\frac{1}{d-1}
(-\tfrac{i}{2}\stackrel{\leftrightarrow}{\bm{\nabla}})^{2}
\sigma^i
\right]
\psi_c.
\end{equation}
In the basis of operators ${\cal O}^i_{An}$ and ${\cal O}^i_{Bn}$
for the spin-triplet case,
the matching conditions (\ref{matching-3}) and (\ref{matching-4}) become
\begin{subequations}
\label{coeffs-a-b}%
\begin{eqnarray}
\label{coeffs-a-b-0}%
i{\cal A}_{\bar{b}c_1}^{i(0)}
&=&
\sum_n a_n^{(0)}\langle 0|{\cal O}^i_{An}|\bar{b}c_1\rangle^{(0)}
+
\sum_n b_n^{(0)}\langle 0|{\cal O}^i_{Bn}|\bar{b}c_1\rangle^{(0)},
\\
\label{coeffs-a-b-1}%
i{\cal A}_{\bar{b}c_1}^{i(1)}-
\left[i{\cal A}_{\bar{b}c_1}^{i(1)}\right]_{\rm NRQCD}
&=&
\sum_n a_n^{(1)}\langle 0|{\cal O}^i_{An}|\bar{b}c_1\rangle^{(0)}
+
\sum_n b_n^{(1)}\langle 0|{\cal O}^i_{Bn}|\bar{b}c_1\rangle^{(0)}
,
\end{eqnarray}
\end{subequations}
where $a_n$ and $b_n$ are the short distance coefficients corresponding
to the operators ${\cal O}^i_{An}$ and ${\cal O}^i_{Bn}$, respectively.
A similar equation holds in the basis ${\cal O}^i_{An}$ and
${\cal O}^i_{Dn}$, where the associated short distance coefficients are
\begin{subequations}
\label{coeffs-s-d}
\begin{eqnarray}
\label{coeffs-s}%
S_n&=&a_n+\frac{1}{d-1}\,b_n,
\\
\label{coeffs-d}
D_n&=&b_n,
\end{eqnarray}
\end{subequations}
where $S_n$ and $D_n$ are the $S$-wave and $D$-wave components of the
short distance coefficient $V_n$, respectively. For more details,
see \cite{Bodwin:2008vp}.

The $\bar{b} c$ matrix elements for the spin-singlet case in
(\ref{matching-3-5}) and (\ref{matching-4-5}) and those for the
spin-triplet case in (\ref{matching-3-1}) and (\ref{matching-4-1})
are calculable perturbatively as
\begin{subequations}
\label{pert-me}%
\begin{eqnarray}
\langle 0|{\cal O}_{n}|\bar{b}c_1\rangle^{(0)}
&=&\bm{q}^{2n}\eta_b^\dagger \xi_c ,
\\
\langle 0|{\cal O}_{An}^i|\bar{b}c_1\rangle^{(0)}
&=&\bm{q}^{2n}\eta_b^\dagger \sigma^i \xi_c ,
\\
\langle 0|{\cal O}_{Bn}^i|\bar{b}c_1\rangle^{(0)}
&=&\bm{q}^{2n-2}q^i\eta^\dagger_b\bm{q}\cdot\bm{\sigma}\xi_c,
\end{eqnarray}
\end{subequations}
where $\xi_c$ and $\eta_b$ are two-component spinors for the charm quark
and the antibottom quark, respectively. In order to maintain consistency
with our calculations in full QCD, we have taken the $\bar{b}c$ states
to have the nonrelativistic normalization and we have suppressed the
factor $\sqrt{N_c}$ that comes from the color trace.

The most general Lorentz covariant forms of
$i {\cal A}^\mu_{5,\bar{b}c_1}$ and $i {\cal A}^\mu_{\bar{b}c_1}$ are
\begin{subequations}
\label{a5mu-amu}
\begin{eqnarray}
\label{a5mu}
i {\cal A}^\mu_{5,\bar{b}c_1} &=& \overline{v}_b(p_{\bar{b}})
(G_{5} \gamma^\mu+ H_{5} q^\mu  + Q_{5} p^\mu )\gamma_5 u_c(p_c),
\\
\label{amu}
i {\cal A}^\mu_{\bar{b}c_1} &=&
\overline{v}_b(p_{\bar{b}}) (G \gamma^\mu
+ H q^\mu
+ Q p^\mu) u_c(p_c),
\end{eqnarray}
\end{subequations}
for the axial vector and vector currents, respectively. Here,
\begin{subequations}
\label{A-Z-L}%
\begin{eqnarray}
G_5&=&\sqrt{Z_bZ_c} (1+\Lambda_5),
\\
G &=&\sqrt{Z_bZ_c} (1+\Lambda),
\end{eqnarray}
\end{subequations}
where $Z_Q$ for $Q=b$ or $c$ is the wavefunction renormalization
factor of the heavy quark $Q$ at order $\alpha_s$ \cite{Braaten:1995ej}:
\begin{equation}
\label{zq}
\sqrt{Z_Q}=1-\frac{\alpha_sC_F}{2\pi}
\left(
\frac{1}{4\epsilon_{\rm UV}}
+\frac{1}{2\epsilon_{\rm IR}}
+\frac{3}{4}\log\frac{4\pi \mu^2 e^{-\gamma_{\rm E}}}{m_Q^2}
+1
\right),
\end{equation}
where $C_F=(N_c^2-1)/(2N_c)=4/3$, $\mu$ is the renormalization
constant, $\gamma_{\rm E}$ is the Euler-Mascheroni constant, and
the subscripts on $1/\epsilon$ indicate the origins of the singularities.
$\Lambda_5$ and $\Lambda$ in (\ref{A-Z-L}) are the multiplicative
corrections to the axial vector and vector vertices, respectively.
Note that the terms proportional to $p^\mu$ survive in
(\ref{a5mu-amu}) because the weak currents are not conserved while
those terms vanish in the electromagnetic current which is conserved.
At order $\alpha_s^0$, the only nonvanishing contributions in
(\ref{a5mu-amu}) are $G^{(0)}_{5} = G^{(0)} = 1$ and all the other
contributions are absent: $H^{(0)}_5=Q^{(0)}_5=H^{(0)}=Q^{(0)} = 0$.
The leading nonvanishing contributions to $H_5$, $Q_5$, $H$, and $Q$
appear from order $\alpha_s$.

Similarly, the nonvanishing components of the NRQCD counterparts to
$i {\cal A}^\mu_{5,\bar{b}c_1}$ and $i {\cal A}^\mu_{\bar{b}c_1}$
in the CM frame of the $\bar{b}c$ pair are
\begin{subequations}
\label{A-NRQCD}
\begin{eqnarray}
i \left[\mathcal{A}_{5,\bar{b}c_1}^0\right]_{\textrm{NRQCD}}
&=&
\bar{v}_b (p_{\bar{b}}) (G_{5,\textrm{NRQCD}} \gamma^0
             + H_{5,\textrm{NRQCD}} q^0
             + Q_{5,\textrm{NRQCD}} p^0) \gamma_5 u_c(p_c),
             \phantom{xxx}
\\
i \left[\mathcal{A}_{\bar{b}c_1}^i\right]_{\textrm{NRQCD}}
&=&
\bar{v}_b (p_{\bar{b}}) (G_{\textrm{NRQCD}} \gamma^i
             + H_{\textrm{NRQCD}} q^i) u_c(p_c),
\end{eqnarray}
\end{subequations}
where we have used $p^i=0$ in the CM frame of the $\bar{b}c$ pair and
\begin{subequations}
\label{GNRQCD}
\begin{eqnarray}
G_{5,\textrm{NRQCD}}
&=&
[\sqrt{Z_b}\,]_\textrm{NRQCD}
[\sqrt{Z_c}\,]_\textrm{NRQCD}
(1+\Lambda_{5,\textrm{NRQCD}}),
\\
G_{\textrm{NRQCD}}
&=&
[\sqrt{Z_b}\,]_\textrm{NRQCD}
[\sqrt{Z_c}\,]_\textrm{NRQCD}
(1+\Lambda_\textrm{NRQCD}).
\end{eqnarray}
\end{subequations}
The heavy quark wavefunction renormalization in NRQCD,
$[\sqrt{Z_Q}\,]_\textrm{NRQCD}$ for $Q=b$ or $c$
at order $\alpha_s$, is given by \cite{Bodwin:2008vp}
\begin{equation}
\label{zq-NRQCD}
\big[\sqrt{Z_Q}\,\big]_{\rm NRQCD}
=1+\frac{\alpha_sC_F}{2\pi}
\left(
\frac{1}{\epsilon_{\rm UV}}
-\frac{1}{\epsilon_{\rm IR}}
\right) .
\end{equation}
By making use of  the Dirac spinors in (\ref{uv-spinors}) and the Dirac
matrices in (\ref{Dirac-matrices}) and (\ref{gamma5}),
we find that the expressions in
(\ref{a5mu-amu}) and (\ref{A-NRQCD}) are reduced into
\begin{subequations}
\label{spinor-reduction}%
\begin{eqnarray}
\label{spinor-reduction-1}
\bar{v}_b (p_{\bar{b}}) \gamma^0\gamma_5 u_c(p_c)
&=& \mathcal{N}_b\mathcal{N}_c
[(E_b+m_b)(E_c+m_c)-\bm{q}^2]
\,
\eta_b^\dagger\xi_c,
\\
\label{spinor-reduction-2}
\bar{v}_b (p_{\bar{b}}) \gamma_5 u_c(p_c)
&=&
-\mathcal{N}_b\mathcal{N}_c
[(E_b+m_b)(E_c+m_c)+\bm{q}^2]
\,
\eta_b^\dagger\xi_c,
\\
\label{spinor-reduction-3}
\bar{v}_b (p_{\bar{b}}) \gamma^i u_c(p_c)
&=&
\mathcal{N}_b\mathcal{N}_c\big\{
[(E_b+m_b)(E_c+m_c)+\bm{q}^2]
\,
\eta_b^\dagger\sigma^i\xi_c
-2q^i\,\eta_b^\dagger\bm{q}\cdot\bm{\sigma}\xi_c\big\},
\phantom{xx}
\\
\label{spinor-reduction-4}
\bar{v}_b (p_{\bar{b}}) u_c(p_c)
&=&
-\mathcal{N}_b\mathcal{N}_c
(E_b+m_b+E_c+m_c)
\,
\eta_b^\dagger\bm{q}\cdot\bm{\sigma}\xi_c,
\end{eqnarray}
\end{subequations}
where the expressions in (\ref{spinor-reduction}) are valid
to all orders in $\bm{q}$ and we have used the identities
\begin{subequations}
\begin{eqnarray}
(\bm{q}\cdot\bm{\sigma})^2&=&\bm{q}^2,
\\
(\bm{q}\cdot\bm{\sigma})\sigma^i(\bm{q}\cdot\bm{\sigma})
&=&2q^i(\bm{q}\cdot\bm{\sigma})-\sigma^i\bm{q}^2,
\end{eqnarray}
\end{subequations}
that derive from (\ref{sigma-delta}) in $d-1$ spatial
dimensions. Note that we do not encounter the products of
Dirac matrices which involve products of three Pauli matrices
of different indices that bring in the contribution in
(\ref{Pauli-reduction}). Therefore, the threshold expansion
(\ref{spinor-reduction}) is free of ambiguities in $d-1$
spatial dimensions. Substituting (\ref{spinor-reduction})
into (\ref{a5mu-amu}), we find that
\begin{subequations}
\label{LOLO5}
\begin{eqnarray}
i {\cal A}^0_{5,\bar{b}c_1} &=&
  \mathcal{N}_b\mathcal{N}_c
  \bigg\{ G_{5}
  \big[(E_b+m_b)(E_c+m_c) - \bm{q}^{2}\big] \nonumber \\
&&\qquad\quad
 - (H_{5} q^0 + Q_{5}
  p^0)\big[(E_b+m_b)(E_c+m_c) + \bm{q}^{2}\big] \bigg\}
  \eta_{b}^\dagger \xi_{c}
  \label{LO5},\\
i {\cal A}^i_{\bar{b}c_1} &=&
  \mathcal{N}_b\mathcal{N}_c
  \bigg\{ G
  \big[(E_b+m_b)(E_c+m_c) + \bm{q}^{2}\big]
  \eta_{b}^\dagger \sigma^{i}
  \xi_{c}
\nonumber \\
&&\qquad\quad
- \big[2 G +(E_b+m_b+E_c+m_c) H\big] q^{i}
  \eta_{b}^\dagger (\bm{\sigma}\cdot\bm{q}) \xi_{c} \bigg\}
  \label{LO}.
\end{eqnarray}
\end{subequations}
Similarly, the NRQCD amplitudes
$[i {\cal A}^0_{5,\bar{b}c_1}]_{\rm NRQCD}$ and
$[i {\cal A}^i_{\bar{b}c_1}]_{\rm NRQCD}$ in (\ref{A-NRQCD})
are the same as
$i {\cal A}^0_{5,\bar{b}c_1}$ and $i {\cal A}^i_{\bar{b}c_1}$
in (\ref{LOLO5}) except that the coefficients $F$ and $F_5$
are replaced with $F_{\rm NRQCD}$ and $F_{5,{\rm NRQCD}}$,
respectively, for $F=G$, $H$, and $Q$.

According to the matching conditions in (\ref{matching-3}),
$[i {\cal A}^{0(0)}_{5,\bar{b}c_1}]_{\rm NRQCD}$ and
$[i {\cal A}^{i(0)}_{\bar{b}c_1}]_{\rm NRQCD}$
are identical to the order-$\alpha_s^0$ full QCD counterparts
in (\ref{LOLO5}).
In order to obtain the short distance coefficients
at order $\alpha_s^0$, we need to expand
$[i {\cal A}^{0(0)}_{5,\bar{b}c_1}]_{\rm NRQCD}=%
i {\cal A}^{0(0)}_{5,\bar{b}c_1}$ and
$[i {\cal A}^{i(0)}_{\bar{b}c_1}]_{\rm NRQCD}$
$=i {\cal A}^{i(0)}_{\bar{b}c_1}$
as linear combinations of the perturbative NRQCD matrix elements in
(\ref{pert-me}). By making use of these order-$\alpha_s^0$ perturbative
matrix elements and the matching conditions in (\ref{matching-3}),
(\ref{coeffs-a-b-0}), and (\ref{coeffs-s-d}), we can obtain
the short distance coefficients at order $\alpha_s^0$:
\begin{subequations}
\label{an0-bn0}%
\begin{eqnarray}
P^{(0)}_n&=&
\left.
\frac{1}{n!}\left(\frac{\partial}{\partial \bm{q}^2}\right)^n
  \mathcal{N}_b\mathcal{N}_c
  \big[(E_b+m_b)(E_c+m_c) - \bm{q}^{2}\big]
\right|_{\bm{q}^2=0},
\\
\label{sn0}
S^{(0)}_n
&=&a_n^{(0)}+\frac{1}{3}\,b_n^{(0)},
\\
\label{dn0}
D^{(0)}_n
&=&b_n^{(0)},
\end{eqnarray}
\end{subequations}
where we have used the fact that $G^{(0)}_5=G^{(0)}=1$ and
$H^{(0)}_5=Q_5^{(0)}=H^{(0)}=Q^{(0)}=0$
and the short distance coefficients $a_n^{(0)}$ and $b_n^{(0)}$
are given by
\begin{subequations}
\label{an0-bn0-2}
\begin{eqnarray}
a_n^{(0)}&=&
\left.
\frac{1}{n!}\left(\frac{\partial}{\partial \bm{q}^2}\right)^n
  \mathcal{N}_b\mathcal{N}_c
  \big[(E_b+m_b)(E_c+m_c) + \bm{q}^{2}\big]
\right|_{\bm{q}^2=0},
\\
b_n^{(0)}&=&
-
\left.
\frac{2}{(n-1)!}\left(\frac{\partial}{\partial \bm{q}^2}\right)^{n-1}
  \mathcal{N}_b\mathcal{N}_c
\right|_{\bm{q}^2=0}.
\end{eqnarray}
\end{subequations}

At order $\alpha_s^1$, the quantities $\Lambda_5$, $\Lambda$, and $Z_Q$
in (\ref{A-Z-L}) contain divergences. The multiplicative vertex
correction factors $\Lambda_5$ and $\Lambda$ have logarithmic
divergences in the UV and IR regions. They also contain  Coulomb
divergence, which is not analytic in the limit $|\bm{q}|\to 0$.
The wavefunction renormalization constant $Z_Q$ has logarithmic
divergences in the UV and IR regions. However, because of the usual
cancellation between the vertex and fermion wavefunction renormalizations,
$G_5^{(1)}$ and $G^{(1)}$ are free of UV divergences and contain only
IR divergences. The quantities $H^{(1)}_{5}$, $Q^{(1)}_{5}$, $H^{(1)}$, and
$Q^{(1)}$, which contribute only from this order,
may have only Coulomb divergences in the limit $|\bm{q}|\to 0$.
Therefore, in the full QCD amplitudes $i {\cal A}^{0(1)}_{5,\bar{b}c_1}$ and
$i {\cal A}^{i(1)}_{\bar{b}c_1}$, UV divergences cancel and
the amplitudes may have singularities only in the IR region,
which are either Coulomb or logarithmic divergences.
Because NRQCD reproduces full QCD in the IR region,
the IR divergences in
 $G^{(1)}_{5,\textrm{NRQCD}}$,
 $H^{(1)}_{5,\textrm{NRQCD}}$,
 $Q^{(1)}_{5,\textrm{NRQCD}}$,
 $G^{(1)}_{\textrm{NRQCD}}$, and
 $H^{(1)}_{\textrm{NRQCD}}$,
cancel those in
 $G_5^{(1)}$,
 $H_5^{(1)}$,
 $Q_5^{(1)}$,
 $G^{(1)}$, and
 $H^{(1)}$, respectively,
to make the following quantities IR finite:
\begin{subequations}
\label{DeltaAB}%
\begin{eqnarray}
\Delta G_5^{(1)}&=&
G_5^{(1)}- G_{5,\rm NRQCD}^{(1)},\\
\Delta H_5^{(1)}&=&
H_5^{(1)}- H_{5,\rm NRQCD}^{(1)},\\
\Delta Q_5^{(1)}&=&
Q_5^{(1)}- Q_{5,\rm NRQCD}^{(1)},\\
\Delta G^{(1)}&=&
G^{(1)}- G_{\rm NRQCD}^{(1)},\\
\Delta H^{(1)}&=&
H^{(1)}- H_{\rm NRQCD}^{(1)}.
\end{eqnarray}
\end{subequations}
Therefore, the right sides of (\ref{matching-4}) which are
determined by the expressions in (\ref{DeltaAB}) are free of
IR sensitivities. Now we can obtain the short distance coefficients at
order $\alpha_s^1$, by making use of the matching conditions
in (\ref{matching-4}), (\ref{coeffs-a-b-1}), and
(\ref{coeffs-s-d}) as
\begin{subequations}
\label{an1-bn1}%
\begin{eqnarray}
P^{(1)}_n&=&
\frac{1}{n!}
\left(\frac{\partial}{\partial \bm{q}^2}\right)^n
  \mathcal{N}_b\mathcal{N}_c
  \bigg\{ \Delta G_{5}^{(1)}
  \big[(E_b+m_b)(E_c+m_c) - \bm{q}^{2}\big]
   \nonumber \\
&&\qquad\quad
 - [\Delta H_{5}^{(1)} q^0 + \Delta Q_{5}^{(1)}
  p^0]\big[(E_b+m_b)(E_c+m_c) + \bm{q}^{2}\big] \bigg\}
\Bigg|_{\bm{q}^2=0},
\\
\label{dm1}
S^{(1)}_n
&=&a_n^{(1)}+\frac{1}{d-1}\,b_n^{(1)},
\\
\label{dm1d}
D^{(1)}_n
&=&b_n^{(1)},
\end{eqnarray}
\end{subequations}
where the short distance coefficients $a_n^{(1)}$ and $b_n^{(1)}$
are given by
\begin{subequations}
\label{an1-bn1-2}
\begin{eqnarray}
a_n^{(1)}&=&
\left.
\frac{1}{n!}\left(\frac{\partial}{\partial \bm{q}^2}\right)^n
  \mathcal{N}_b\mathcal{N}_c
 \Delta G^{(1)}
  \big[(E_b+m_b)(E_c+m_c) + \bm{q}^{2}\big]
\right|_{\bm{q}^2=0},
\\
b_n^{(1)}&=&
-
\left.
\frac{1}{(n-1)!}\left(\frac{\partial}{\partial \bm{q}^2}\right)^{n-1}
\!\!\!\!\!  \mathcal{N}_b\mathcal{N}_c
\big[2 \Delta G^{(1)}
 +(E_b+m_b+E_c+m_c) \Delta H^{(1)}\big]
\right|_{\bm{q}^2=0}.
\nonumber\\
\end{eqnarray}
\end{subequations}
Although we have completely removed the IR singularities in the short distance
coefficients (\ref{an1-bn1}) and (\ref{an1-bn1-2}), they have logarithmic
UV divergences that are originated from the one-loop NRQCD matrix elements in
$G_{5,\rm NRQCD}^{(1)}$ and $G_{\rm NRQCD}^{(1)}$. As stated before,
the quantities  $\Delta H_5^{(1)}$, $\Delta Q_5^{(1)}$, and $\Delta H^{(1)}$
are free of UV divergences as well as IR divergences. We renormalize
$\Delta G_{5}^{(1)}$, $\Delta G^{(1)}$, and the
short distance coefficients in (\ref{an1-bn1}) and (\ref{an1-bn1-2})
according to the $\overline{\rm MS}$ scheme to find that
\begin{equation}
\label{an1-bn1-MSbar}%
\big[f^{\,(1)}_n\big]_{\overline{\rm MS}}=
f^{\,(1)}_n\big|_{
\Delta F^{(1)}\to\, \Delta F^{(1)}_{\overline{\rm MS}}
},
\end{equation}
where $F=G_5$ or $G$ and $f=P$, $a$, $b$, $S$, and $D$.
In deriving the expression for $\big[f_n^{\,(1)}\big]_{\overline{\rm MS}}$
in (\ref{an1-bn1-MSbar}), we have used the fact that,
in minimal subtraction, one removes the $1/\epsilon$ pole times the
order-$\alpha_s^0$ $d$-dimensional matrix element. Hence, a term proportional
to $(d-1)^{-1}\epsilon^{-1}$ is subtracted in (\ref{dm1}) in carrying out
the renormalization.

\section{QCD corrections\label{sec:QCD}}
In this section, we compute the one-loop QCD corrections to the axial vector
and vector parts of the charged weak current
$i\mathcal{A}_{5,\bar{b}c_1}^{0(1)}$ and $i\mathcal{A}_{\bar{b}c_1}^{0(1)}$,
respectively. The order-$\alpha_s$ QCD corrections are composed
of the vertex corrections and the wavefunction renormalization
contributions. In the Feynman gauge, the vertex correction contributions
to $i\mathcal{A}_{5,\bar{b}c_1}^{0(1)}$ and
$i\mathcal{A}_{\bar{b}c_1}^{0(1)}$ are given by
\begin{subequations}
\label{mu5mu}
\begin{eqnarray}
\label{mu5}
\Lambda^\mu_5 &=& - i g_s^2 C_F \int_k
\frac{\overline{v} (p_{\bar{b}}) \gamma_\alpha
(-\!\not{\!p_{\bar{b}}}\, +\! \not{\!k} + m_b) \gamma^\mu \gamma_5
(\not{\!p_c} +\!\not{\!k} + m_c )
\gamma^\alpha u (p_c)}{D_0D_1D_2},
\\
\label{mu}
\Lambda^\mu &=& - i g_s^2 C_F \int_k
\frac{
\overline{v} (p_{\bar{b}}) \gamma_\alpha
(-\!\not{\!p_{\bar{b}}}\, +\! \not{\!k} + m_b) \gamma^\mu
(\not{\!p_c} + \!\not{\!k} + m_c )
\gamma^\alpha u (p_c)}{D_0 D_1 D_2} ,
\end{eqnarray}
\end{subequations}
where $g_s^2=4\pi \alpha_s$ is the strong coupling
and the symbol $\int_k$ and the denominator factors $D_i$'s
are defined by
\begin{subequations}
\label{intk-di}
\begin{eqnarray}
\int_k&\equiv&\mu^{2\epsilon}\int \frac{d^d k}{(2\pi)^d},
\\
D_{0} &=& k^2 + i \varepsilon,
\\
D_{1} &=& k^2 - 2 k \cdot p_{\bar{b}} + i\varepsilon,
\\
D_{2} &=& k^2 + 2 k \cdot p_c + i \varepsilon.
\end{eqnarray}
\end{subequations}
Here, $\mu$ is the renormalization scale and we have taken the gluon
momentum as the loop momentum $k$.
By making use of the anticommutation relation of $\gamma_5$ and $\gamma^\mu$
we rearrange each term of the numerator in (\ref{mu5mu}) as a linear
combination of terms
$\gamma_\alpha \Gamma_i\gamma^\alpha\gamma_5$
or
$\gamma_\alpha \Gamma_i\gamma^\alpha$,
where $\Gamma_i$ is a product of three or less Dirac matrices.
Applying the anticommutation relation
(\ref{anticommutation-relation}) summed over the $d$-dimensional
index $\alpha$ and making use of the on-shell conditions
\begin{subequations}
\label{Dirac-equation}
\begin{eqnarray}
/\!\!\!p_{c}u(p_{c})&=&m_cu(p_{c}),
\\
/\!\!\!p_{\bar{b}}v(p_{\bar{b}})&=&-m_bv(p_{\bar{b}}),
\end{eqnarray}
\end{subequations}
we can reduce the expressions in (\ref{mu5mu}) as
\begin{subequations}
\label{QCD-lambda}
\begin{eqnarray}
\label{qcdlmu}
\Lambda^\mu_{5} &=&-ig_s^2 C_F
\int_{k}\frac{\overline{v}(p_{\bar{b}}) \Gamma^\mu(k,p,q,m_b,-m_c)
\gamma_5 u(p_c)}{D_0D_1D_2}
,
\\
\label{qcdmu}
\Lambda^\mu &=&-ig_s^2 C_F
\int_{k}\frac{\overline{v}(p_{\bar{b}}) \Gamma^\mu(k,p,q,m_b,m_c) u(p_c)}
{D_0D_1D_2},
\end{eqnarray}
\end{subequations}
where $\Gamma^\mu(k,p,q,m_b,m_c)$ is defined by
\begin{eqnarray}
\label{GAMMA-MU}
\Gamma^\mu(k,p,q,m_b,m_c)&=&\big[(d-2)k^2
-  2(4p^2-m_b^2-m_c^2) + 8 k\cdot q\big]\gamma^\mu
\nonumber\\&&
+ 2 (m_b  \gamma^\mu /\!\!\!k +m_c /\!\!\!k \gamma^\mu)
+2(2-d)k^\mu /\!\!\!k
- 8 q^\mu /\!\!\!k.
\end{eqnarray}
By making use of the standard reduction methods for the tensor
loop integrals, we can express all of the loop-momentum dependence
in terms of $p_{\bar{b}}$ and $p_c$, which are linear
combinations of $p$ and $q$. Because $p\cdot q\neq0$, the reduction
formulas for the tensor integrals of the $S$-wave $\bar{b}c_1$ decay are
slightly more complicated than those for the spin-triplet $S$-wave
$Q\bar{Q}_1$ decay in \cite{Bodwin:2008vp} where $p \cdot q = 0$.
In appendix~\ref{appendix:tensor} we list the formulas for the tensor
reduction. Once we apply the equations of motion in (\ref{Dirac-equation}),
$\Gamma^\mu$'s in (\ref{QCD-lambda}) are reduced into a linear combination
of $\gamma^\mu$, $p^\mu\mathbbm{1}$, and $q^\mu\mathbbm{1}$ as
\begin{subequations}
\label{l5mu-lmu}
\begin{eqnarray}
\label{l5mu}
\Lambda^\mu_5 &=& - i g_s^2 C_F \overline{v} (p_{\bar{b}})
 \big[A_1(-m_c)\gamma^\mu
     +A_2(-m_c)p^\mu
     +A_3(-m_c)q^\mu
\big] \gamma_5 u(p_c),
\\
\label{lmu}
\Lambda^\mu &=& - i g_s^2 C_F \overline{v} (p_{\bar{b}})
 \big[A_1(m_c)\gamma^\mu
     +A_2(m_c)p^\mu
     +A_3(m_c)q^\mu
\big] u(p_c).
\end{eqnarray}
\end{subequations}
As mentioned in \cite{Hwang:1999fc}, the Lorentz scalar coefficients
$A_i$'s in (\ref{l5mu}) and (\ref{lmu}) are the same except for the
replacement $A_i(-m_c) \leftrightarrow A_i(m_c)$ and $A_i(m_c)$'s are defined by
\begin{subequations}
\label{Ai}
\begin{eqnarray}
A_1(m_c)&=&(d-2)(J_1-2J_4) - 4p_{\bar{b}} \cdot p_c J_2  + 4 J_3
+ \frac{(m_b-m_c)^2J_3+(m_b^2-m_c^2)J_5}
       {p_{\bar{b}} \cdot p_c + m_b m_c},\phantom{xxxxx}
\\
A_2(m_c)&=&
2 \bigg[\frac{(m_b-m_c)J_3+(m_b+m_c)J_5}
             {p_{\bar{b}} \cdot p_c + m_b m_c}
+\frac{(2-d)J_7(m_c)}{2 p^2 q^2}\bigg],
\\
A_3(m_c)&=&  2 \bigg[
\frac{(m_b+m_c)J_3+(m_b-m_c)J_5}
             {p_{\bar{b}} \cdot p_c - m_b m_c}
+\frac{(2-d)J_6(m_c)}{2q^2}\bigg].
\end{eqnarray}
\end{subequations}
Here, the $J_i$'s are scalar integrals which are defined
and evaluated in appendix~\ref{appendix:scalarint}.
Except for the two scalar integrals $J_6$ and $J_7$,
the other scalar integrals $J_i$ for $i\le 5$ are even functions
of $m_c$ and, therefore, we suppress the arguments of
$J_i$ for $i\le 5$. The expressions in (\ref{l5mu-lmu})
are new, in which the relativistic corrections to all orders in
$\bm{q}^{2}$ are included.

According to (\ref{a5mu}) and (\ref{A-Z-L}),
the vertex corrections can be parametrized as
$\Lambda^\mu_5=\bar{v}(p_{\bar{b}})
(\Lambda_5\gamma^\mu +H_5p^\mu +Q_5 q^\mu)\gamma_5 u(p_c)$
and $\Lambda^\mu=\bar{v}(p_{\bar{b}})
(\Lambda\gamma^\mu +Hp^\mu +Q q^\mu) u(p_c)$.
Therefore, the multiplicative factors
$\Lambda_5$ and $\Lambda$ for the vertex corrections can be determined
by $A_1(\mp m_c)$, which are the coefficients of $\gamma^\mu$ in
(\ref{l5mu-lmu}). In similar ways, $H_5$ and $H$ are determined
by $A_2(\mp m_c)$ and $Q_5$ and $Q$ are determined by $A_3(\mp m_c)$,
respectively. Substituting the values for the scalar integrals $J_i$'s
evaluated in appendix~\ref{appendix:scalarint} into $A_1(\mp m_c)$
in (\ref{Ai}) and then substituting $A_1(\mp m_c)$ into
(\ref{l5mu-lmu}), we obtain the multiplicative factors
for the vertex corrections as $\Lambda_5=-ig_s^2C_F A_1(-m_c)$ and
$\Lambda=-ig_s^2C_F A_1(m_c)$. The results are
\begin{subequations}
\label{la5la}
\begin{eqnarray}
\Lambda_5 &=&
\frac{\alpha_s C_F}{4\pi}\Bigg\{\frac{1}{\epsilon_{\rm UV}}
+ \frac{1}{2}\log
\frac{(4\pi\mu^2 e^{-\gamma_{\rm E}})^2}{m_b^2 m_c^2}
+\bigg[
  \frac{p_{\bar{b}}\cdot p_c}{2p^2}
  \bigg(\frac{1}{\epsilon_{\rm IR}}+\frac{1}{2}\log
\frac{(4\pi\mu^2 e^{-\gamma_{\rm E}})^2}{m_b^2 m_c^2}\bigg)
\nonumber\\
&&+\,
\delta^2\left(6 + \frac{2(m_b+m_c)^2}{p_{\bar{b}} \cdot p_c - m_b m_c}\right)
\bigg]
\left[L_1(\delta,e_c) -  \frac{i\pi}{\delta}\right]
-\frac{p_{\bar{b}}\cdot p_c}{p^2}\, K(\delta,e_c)
-\frac{1}{8p^2}
   \big[m_b^2-m_c^2
\nonumber\\
&&
+2p_{\bar{b}}\cdot p_c \,L_2(\delta,e_c)
   \big]\log\frac{m_c^2}{m_b^2}
  +\frac{p_{\bar{b}}\cdot p_c}{2p^2}
\bigg[\frac{\pi^2}{\delta}
-
  \frac{i\pi}{2\delta} \log \frac{p^4 m_b^2 m_c^2}
  {[(p_{\bar{b}} \cdot p_c)^2-m_b^2 m_c^2]^2} \bigg]\Bigg\},
\\
\Lambda &=&\Lambda_5|_{m_c\to -m_c},
\end{eqnarray}
\end{subequations}
where
$\Lambda_5|_{m_c\to -m_c}$ represents the expression that
can be obtained from $\Lambda_5$ after replacing $m_c$ with $-m_c$
and the variables $\delta$ and $e_c$ are defined in (\ref{def-ab}).
The functions $L_1(\delta,e_c)$, $L_2(\delta,e_c)$, and $K(\delta,e_c)$ are defined by
\begin{subequations}
\label{def-LK}
\begin{eqnarray}
L_1(\delta,e_c) &=& \frac{1}{2\delta}
\log \frac{(e_c+\delta)}{(e_c-\delta)}
     \frac{(1-e_c+\delta)}{(1-e_c-\delta)}\nonumber\\
&=&\frac{(m_b+m_c)^2}{m_b m_c}
  -\frac{(m_b+m_c)^2 \left(m_b^2-4m_c m_b+m_c^2\right)}
        {6 m_b^3 m_c^3}\,{\bm q}^2
 +{O}({\bm q}^4),
\\
L_2(\delta,e_c) &=& \frac{1}{2\delta}
\log \frac{(e_c+\delta)}{(e_c-\delta)}
     \frac{(1-e_c-\delta)}{(1-e_c+\delta)}\nonumber\\
&=&\frac{m_b^2-m_c^2}{m_bm_c}-\frac{(m_b+m_c) (m_b-m_c)^3}
        {6 m_b^3m_c^3}\,{\bm q}^2
 +{O}({\bm q}^4),
\\
\label{def-K}
K(\delta,e_c) &=& \frac{1}{4\delta }\left[
  {\rm Sp}\!\left(\frac{2\delta}{e_c+\delta}\right)
- {\rm Sp}\!\left(\frac{-2\delta}{e_c-\delta}\right)
+ {\rm Sp}\!\left(\frac{2\delta}{1-e_c+\delta}\right)
- {\rm Sp}\!\left(\frac{-2\delta}{1-e_c-\delta}\right)
 \right]\nonumber\\
 &=&\frac{(m_b+m_c)^2}{m_b m_c}-\frac{(m_b+m_c)^2 \left(m_b^2-10
   m_c m_b+m_c^2\right)}{18 m_b^3 m_c^3}\,{\bm q}^2
   +{O}({\bm q}^4).\phantom{xxxxx}
\end{eqnarray}
\end{subequations}
We have listed the first two leading terms in the nonrelativistic
expansions of these functions. According to (\ref{def-LK}),
the functions $L_1(\delta,e_c)$, $L_2(\delta,e_c)$, and
$K(\delta,e_c)$ are of order 1 as $|\bm{q}|\to 0$.
The function ${\rm Sp}(x)$ in (\ref{def-K}) is the Spence
function, which is defined by
\begin{equation}
{\rm Sp}(x) = \int_{x}^0 \frac{\log (1-t)}{t}dt.
\end{equation}

Now we evaluate $G_5$ and $G$.
Substituting $\Lambda_5$ and $\Lambda$ in (\ref{la5la})
and the heavy quark wavefunction renormalization constant $Z_Q$
for $Q=b$ and $c$ in (\ref{zq}) into (\ref{A-Z-L}), we obtain
\begin{subequations}
\label{G-final}
\begin{eqnarray}
G_{5} &=&
1 + \frac{\alpha_s C_F}{4\pi} \Bigg\{
  \bigg[\frac{p_{\bar{b}}\cdot p_c}{2p^2}
  \left(L_1(\delta,e_c) - \frac{i\pi}{\delta}\right) -2\bigg]
  \bigg(\frac{1}{\epsilon_{\rm IR}}
  + \frac{1}{2}\log
\frac{(4\pi\mu^2 e^{-\gamma_{\rm E}})^2}{m_b^2 m_c^2}
  \bigg)
  \nonumber \\
&&+
  \bigg[6 + \frac{2 (m_b+m_c)^2}{p_{\bar{b}} \cdot p_c - m_b m_c}
  \bigg] \delta^2
  \bigg[L_1(\delta,e_c) - \frac{i\pi}{\delta}\bigg]
  - \frac{p_{\bar{b}}\cdot p_c}{p^2} \, K(\delta,e_c)
  - 4
  -\frac{1}{8p^2}\big[m_b^2-m_c^2
  \nonumber \\&&
  +2p_{\bar{b}}\cdot p_c\, L_2(\delta,e_c)\big]
  \log \frac{m_c^2}{m_b^2}
  +\frac{p_{\bar{b}}\cdot p_c}{2p^2}
   \bigg[\frac{\pi^2}{\delta} -\frac{i\pi}{2\delta}
   \log \frac{p^4 m_b^2 m_c^2}
  {[(p_{\bar{b}} \cdot p_c)^2-m_b^2 m_c^2]^2}
  \bigg] \Bigg\},
\\
G&=&G_{5}|_{m_c\to -m_c}.
\end{eqnarray}
\end{subequations}
As we have expected, $G_5$ and $G$ are free of UV divergences
while they have Coulomb and logarithmic divergences
in the IR region.

Next we substitute the values for the scalar integrals $J_i$'s
evaluated in appendix~\ref{appendix:scalarint} into $A_2(\mp m_c)$ and
$A_3(\mp m_c)$ in (\ref{Ai}). Then we can determine
$H_5=-ig_s^2C_F A_2(-m_c)$, $Q_5=-ig_s^2C_F A_3(-m_c)$,
$H=-ig_s^2C_F A_2(m_c)$, and $Q=-ig_s^2C_F A_3(m_c)$.
The results are
\begin{subequations}
\label{HQ-final}
\begin{eqnarray}
H_{5} &=&\frac{\alpha_s C_F}{4\pi}\;
\frac{2(m_b-m_c)}{p_{\bar{b}} \cdot p_c + m_b m_c}
  \bigg\{ \delta^2 \bigg[L_1(\delta,e_c)-\frac{i\pi}{\delta}\bigg]
   + r_-\log \frac{m_c^2}{m_b^2}
  \bigg\},
\\
Q_{5} &=&\frac{\alpha_s C_F}{4\pi}(m_b+m_c)
\Bigg(
\frac{4}{p_{\bar{b}} \cdot p_c - m_b m_c}
  \bigg\{ \delta^2\bigg[L_1(\delta,e_c)-\frac{i\pi}{\delta}
                      \bigg]
   +r_+\log\frac{m_c^2}{m_b^2}\bigg\}
\nonumber\\&&
  -\frac{(m_b-m_c)^2}{2p^2(p_{\bar{b}}\cdot p_c + m_b m_c)}
  \bigg\{ \delta^2
  \bigg[L_1(\delta,e_c)- \frac{i\pi}{\delta}\bigg]
  + r_-\log  \frac{m_c^2}{m_b^2} \bigg\}
  \nonumber \\
&& +  \frac{1}{p^2}
\bigg\{\delta^2\bigg[L_1(\delta,e_c)- \frac{i\pi}{\delta}\bigg]
- 1+ r_+
 \log
  \frac{m_c^2}{m_b^2}  \bigg\}
  \Bigg),\phantom{xxxxx}
\\
H&=&
H_{5}|_{m_c\to -m_c},
\\
Q&=&
Q_{5}|_{m_c\to -m_c},
\end{eqnarray}
\end{subequations}
where $r_\pm$ is defined by
\begin{equation}
r_\pm=\frac{m_b\mp m_c}{4(m_b\pm m_c)}
\bigg[1-\frac{(m_b\pm m_c)^2}{4p^2}\bigg].
\end{equation}
The variable $r_-$ is of order 1 and the variable $r_+$ is of
order $\bm{q}^2$, which are both finite in the limit $|\bm{q}|\to 0$.
The real parts of the quantities in (\ref{HQ-final}) are finite.
According to (\ref{pbpc}) and (\ref{def-ab}), $\textrm{Im}\,H_{5}$
is of order $|\bm{q}|$ and, therefore, finite and the leading contribution
to $\textrm{Im}\, Q_{5,{\rm NRQCD}}$ is of order $1/|\bm{q}|$ and
acquires Coulomb divergence. In the case of the vector counterparts,
the leading contributions to $\textrm{Im}\,H$ and
$\textrm{Im}\,Q$ are both Coulomb divergent.
\section{NRQCD corrections\label{sec:NRQCD}}
In this section, we compute the NRQCD amplitudes of order $\alpha_s^1$.
As shown in section~\ref{sec:QCD}, the one-loop QCD corrections
$i\mathcal{A}^{0(1)}_{5,\bar{b}c_1}$ and
$i\mathcal{A}^{i(1)}_{\bar{b}c_1}$ contain Coulomb and logarithmic divergences
in the IR regions. In order to determine the short distance coefficients
$P_n^{(1)}$ and $V_n^{(1)}$ in (\ref{matching-4}),
which are insensitive to the long distance interactions, we remove
those divergences based on the fact that NRQCD amplitudes
must reproduce the corresponding full QCD amplitudes in the IR regions
because NRQCD is a low energy effective field theory of QCD.
We shall find that the divergences of $i\mathcal{A}^{0(1)}_{5,\bar{b}c_1}$ and
$i\mathcal{A}^{i(1)}_{\bar{b}c_1}$ are identified as the one-loop
corrections to the perturbative NRQCD matrix elements
$\langle 0|\mathcal{O}_n|\bar{b}c_1\rangle^{(1)}$ and
$\langle 0|\mathcal{O}^i_n|\bar{b}c_1\rangle^{(1)}$ in (\ref{matching-2}).

Instead of following the direct NRQCD approach, we compute NRQCD quantities
from the full QCD expressions $i\mathcal{A}^0_{5,\bar{b}c_1}$ and
$i\mathcal{A}^i_{\bar{b}c_1}$ based on the method in \cite{Bodwin:2008vp}.
First we carry out the integration over the temporal component $k^0$ of the
loop integral and, then, expand the integrand in powers of $\bm{q}/m_Q$ and
$\bm{k}/m_Q$, where $\bm{k}$ is the spatial component of the loop momentum.
We regularize divergent integrals dimensionally and drop scaleless
power-divergent integrals. The only nonvanishing divergent
contributions are, then, either logarithmic or Coulomb divergent.
As in \cite{Bodwin:2008vp}, we use a special notation for this
prescription for the loop integration as
\raisebox{-0.6ex}{${}^\mathcal{N}$}\hspace{-1.7ex}$\int_{k}$.
Once $k^0$ integral has been evaluated by contour integration, then the
remaining integral is denoted as
\begin{equation}
\mathcal{N}\!\!\!\!\!\!\!\int_{\bm{k}}\equiv
 \mu^{2\epsilon}\mathcal{N}\!\!\!\!\!\!\!\int\frac{d^{d-1}k}{(2\pi)^{d-1}},
\end{equation}
where \raisebox{-0.6ex}{${}^\mathcal{N}$}\hspace{-1.7ex}$\int_{\bm{k}}$
indicates that the integrand of the spatial loop variable must be expanded
in powers of $\bm{q}/m_Q$ and $\bm{k}/m_Q$ and then regulated dimensionally
in $d-1$ spatial dimensions.

To evaluate the vertex corrections $\Lambda^\mu_{5,{\rm NRQCD}}$
and $\Lambda^\mu_{\rm NRQCD}$ in NRQCD, we begin with the full QCD
expressions in (\ref{QCD-lambda}) by replacing the loop integrals
$\int_k$ with
\raisebox{-0.6ex}{${}^\mathcal{N}$}\hspace{-1.7ex}$\int_{k}$:
\begin{subequations}
\label{NRQCD-lambda0}
\begin{eqnarray}
\Lambda^\mu_{5,{\rm NRQCD}} &=&-ig_s^2 C_F
\mathcal{N}\!\!\!\!\!\!\!\int_{k}\frac{
\overline{v}(p_{\bar{b}}) \Gamma^\mu(k,p,q,m_b,-m_c)
\gamma_5 u(p_c)}{D_0D_1D_2}
, \label{nrqcdlmu0}
\\
\Lambda^\mu_{\rm NRQCD} &=&-ig_s^2 C_F
\mathcal{N}\!\!\!\!\!\!\!\int_{k}\frac{
\overline{v}(p_{\bar{b}}) \Gamma^\mu(k,p,q,m_b,m_c)
u(p_c)}{D_0D_1D_2},
 \label{nrqcdmu0}
\end{eqnarray}
\end{subequations}
where $\Gamma^\mu(k,p,q,m_b,m_c)$ is defined in (\ref{GAMMA-MU}).
In the NRQCD case, we omit the tensor reduction and
directly evaluate the integrals
 \raisebox{-0.6ex}{${}^\mathcal{N}$}\hspace{-1.7ex}$\int_{k}$
because the tensor reduction does not simplify
the intermediate steps of calculation considerably.
Then the expressions in (\ref{NRQCD-lambda0}) becomes
\begin{subequations}
\label{NRQCD-lambda}
\begin{eqnarray}
\Lambda^0_{5,{\rm NRQCD}} &=&-ig_s^2 C_F \overline{v}(p_{\bar{b}})
\bigg\{\big[(d-2)S_1 -
  2(4p^2-m_b^2-m_c^2)S_2 + 8 q_\nu S_3^\nu\big]\gamma^0
   \nonumber \\
&&
+2 (m_b\gamma^0 \gamma_\nu -m_c \gamma_\nu\gamma^0)S_3^\nu  +
  2(2-d)S_4^{0\nu}\gamma_\nu  - 8 q^0 S_3^\nu \gamma_\nu
  \bigg\} \gamma_5 u(p_c), \label{nrqcdlmu}
\\
\Lambda^i_{\rm NRQCD} &=&
-ig_s^2 C_F \overline{v}(p_{\bar{b}})
\bigg\{\big[(d-2)S_1 -
  2(4p^2-m_b^2-m_c^2)S_2 + 8 q_\nu S_3^\nu\big]\gamma^i
   \nonumber \\
&&
+2 (m_b\gamma^i \gamma_\nu +m_c \gamma_\nu\gamma^i)S_3^\nu  +
  2(2-d)S_4^{i\nu}\gamma_\nu  - 8 q^i S_3^\nu \gamma_\nu
  \bigg\}  u(p_c),
\label{nrqcdmu}
\end{eqnarray}
\end{subequations}
where the loop integrals $S_1$, $S_2$, $S_3^\mu$, and $S_4^{\mu\nu}$
are defined by
\begin{subequations}
\label{S-integrals}
\begin{eqnarray}
S_1&=&\mathcal{N}\!\!\!\!\!\!\!\int_{k}\frac{1}{D_1D_2},
\\
S_2&=&\mathcal{N}\!\!\!\!\!\!\!\int_{k}\frac{1}{D_0D_1D_2},
\\
S_3^\mu&=&\mathcal{N}\!\!\!\!\!\!\!\int_{k}\frac{k^\mu}{D_0D_1D_2},
\\
S_4^{\mu\nu}
&=&\mathcal{N}\!\!\!\!\!\!\!\int_{k}\frac{k^\mu k^\nu}{D_0D_1D_2}.
\end{eqnarray}
\end{subequations}

Next we carry out the $k^0$ integrals in (\ref{NRQCD-lambda}) by
contour integration. In order to identify the residues of the $k^0$ integral,
we express the denominator factors of the integrands in
(\ref{S-integrals}) in the following form:
\begin{subequations}
\label{D0-NRQCD}
\begin{eqnarray}
D_{0} &=& (k^0)^2-\bm{k}^2 + i \varepsilon
= (k^0 - |\bm{k}| + i \varepsilon)
  (k^0 + |\bm{k}| - i \varepsilon),
\\
D_{1} &=& (k^0-E_b)^2-\Delta_b^2+ i \varepsilon
= (k^0 + \Delta_b - E_b  - i \varepsilon)
  (k^0 - \Delta_b - E_b  + i \varepsilon),
\\
D_{2} &=& (k^0+E_c)^2-\Delta_c^2+ i \varepsilon
= (k^0 + \Delta_c + E_c  - i \varepsilon)
  (k^0 - \Delta_c + E_c  + i \varepsilon),
\end{eqnarray}
\end{subequations}
where $\Delta_Q$ for $Q=c$ or $b$ are defined by
\begin{equation}
\label{Delta-Q}
\Delta_Q =\sqrt{ m_Q^2 + (\bm{k}+\bm{q})^2}.
\end{equation}

The resultant integrands of the $(d-1)$-dimensional integrals
\raisebox{-0.6ex}{${}^\mathcal{N}$}\hspace{-1.7ex}$\int_{\bm{k}}$
are then expanded in powers of $\bm{q}/m_Q$ and
$\bm{k}/m_Q$ for $Q=b$ or $c$. The following relations are useful
in these expansions:
\begin{subequations}
\label{useful-formulas}
\begin{eqnarray}
\Delta_Q-E_Q &=& \frac{{\bm k}^2+2 {\bm k}\cdot {\bm q}}{\Delta_Q+E_Q},
\\
\Delta_b^2-\Delta_c^2&=&m_b^2-m_c^2,
\\
\Delta_Q^2-(E_Q\pm |{\bm k}|)^2&=&
\mp2|{\bm k}|(E_Q\mp {\bm q}\cdot \hat{{\bm k}}).
\end{eqnarray}
\end{subequations}
It is evident from (\ref{useful-formulas}) that the factors such as
$1/(\Delta_Q-E_Q)$ and $1/[\Delta_Q^2-(E_Q\pm |{\bm k}|)^2]$ may give rise
to IR singularities. In this step, scaleless integrals that are power divergent
in the UV regions are neglected under dimensional regularization.
The nonvanishing elementary integrals $n_0$, $n_1$, $n_2$, and $n_3$ that
survive in this step are evaluated in appendix~\ref{appendix:NRscalarint}.
Nonvanishing scaleless integrals are logarithmically divergent,
which are proportional to $n_0$ defined in (\ref{n0}). The integrals
$n_1$, $n_2$, and $n_3$ defined in (\ref{n123}) have scale dependencies
on $|\bm{q}|$. Eventually, all of the loop integrals in (\ref{S-integrals})
are decomposed into linear combinations of these elementary integrals.
The resultant values for the integrals (\ref{S-integrals}) are given in
(\ref{s1-final}), (\ref{s2-final}), (\ref{s3-final}), and (\ref{s4-final})
of appendix~\ref{appendix:NRscalarint}. Substituting these values to
(\ref{NRQCD-lambda}), we find the multiplicative vertex correction factors
in NRQCD as
\begin{subequations}
\label{Lambda-NRQCD-f}
\begin{eqnarray}
\Lambda_{5,{\rm NRQCD}} &=& \frac{\alpha_s C_F}{4\pi}
\frac{p_{\bar{b}}\cdot p_c}{2p^2}
  \bigg\{
  \bigg(\frac{1}{\epsilon_{\rm IR}} - \frac{1}{\epsilon_{\rm UV}}
  \bigg) L_1(\delta,e_c)+\frac{\pi^2}{\delta}
-\frac{i\pi}{\delta}\bigg[\frac{1}{\epsilon_{\rm IR}}
\nonumber \\&&
+ \log
  \bigg(\frac{\pi\mu^2e^{-\gamma_{\rm E}}}{\bm{q}^2}\bigg)
  +\bigg(\!6 +\frac{2(m_b+m_c)^2}
                 {p_{\bar{b}} \cdot p_c - m_bm_c}
   \bigg)\frac{\bm{q}^2}{2p_{\bar{b}} \cdot p_c} \bigg] \bigg\},
\\
\Lambda_{\rm NRQCD} &=& {[\Lambda_5]}_{\rm NRQCD}\big|_{m_c\to -m_c}.
\end{eqnarray}
\end{subequations}
Substituting the multiplicative vertex correction factors
in (\ref{Lambda-NRQCD-f}) and the
heavy quark wavefunction renormalization factor
in (\ref{zq-NRQCD}) into (\ref{GNRQCD}),
we obtain $G_{5,{\rm NRQCD}}$ and $G_{\rm NRQCD}$ as
\begin{subequations}
\label{G-NRQCD-final}
\begin{eqnarray}
G_{5,{\rm NRQCD}}
&=& 1 + \frac{\alpha_s C_F}{4\pi}
\frac{p_{\bar{b}}\cdot p_c}{2p^2}
  \bigg\{
  \bigg(\frac{1}{\epsilon_{\rm IR}} - \frac{1}{\epsilon_{\rm UV}}
  \bigg) \bigg( L_1(\delta,e_c)-\frac{4p^2}{p_{\bar{b}} \cdot p_c}\bigg)
  +\frac{\pi^2}{\delta}
\nonumber \\&&\!\!\!\!\!\!\!
-\frac{i\pi}{\delta}\bigg[\frac{1}{\epsilon_{\rm IR}}
+ \log\!
  \bigg(\frac{\pi\mu^2e^{-\gamma_{\rm E}}}{\bm{q}^2}\bigg)
  \!+\!\bigg(\!6 +\frac{2(m_b+m_c)^2}
                 {p_{\bar{b}} \cdot p_c - m_bm_c}
   \bigg)\frac{\bm{q}^2}{2p_{\bar{b}} \cdot p_c} \bigg] \bigg\},
\phantom{xxxx}
\\
G_{\rm NRQCD} &=&G_{5,{\rm NRQCD}}  |_{m_c\to -m_c}.
\end{eqnarray}
\end{subequations}
As was expected, the logarithmic and Coulomb divergences in the IR regions
of $G_5$ and $G$ in (\ref{G-final}) are reproduced in
$G_{5,{\rm NRQCD}}$ and $G_{\rm NRQCD}$ in (\ref{G-NRQCD-final}),
respectively. We notice that unlike $G_5$ and $G$, $G_{5,{\rm NRQCD}}$ and
$G_{\rm NRQCD}$ contain logarithmic UV divergences.
In a similar way, the remaining NRQCD correction factors are obtained as
\begin{subequations}
\label{HQ-NRQCD-final}
\begin{eqnarray}
H_{5,{\rm NRQCD}} &=& -\frac{\alpha_s C_F}{4\pi}
i\pi \delta\;
\bigg[
\frac{2(m_b-m_c)}{p_{\bar{b}} \cdot p_c + m_b m_c}
\bigg],
\\
Q_{5,{\rm NRQCD}} &=& -\frac{\alpha_s C_F}{4\pi}
\frac{i\pi \delta(m_b+m_c)}{p^2}
\bigg\{
1
- \bigg[
\frac{(m_b-m_c)^2}{2(p_{\bar{b}}  \cdot p_c + m_b m_c)}
\nonumber\\
&&
-\frac{4p^2}{p_{\bar{b}} \cdot p_c -  m_b m_c}
\bigg]
\bigg\},
\\
{H}_{\rm NRQCD} &=&H_{5,{\rm NRQCD}}\big|_{m_c\to -m_c},
\\
{Q}_{\rm NRQCD} &=&Q_{5,{\rm NRQCD}}\big|_{m_c\to -m_c}.
\end{eqnarray}
\end{subequations}
All of the quantities in (\ref{HQ-NRQCD-final}) are pure imaginary.
According to (\ref{def-ab}) and (\ref{pbpc}), $H_{5,{\rm NRQCD}}$
is of order $|\bm{q}|$ and, therefore, finite and the leading contribution
to $Q_{5,{\rm NRQCD}}$ is of order $1/|\bm{q}|$ in the limit $|\bm{q}|\to 0$
and acquires Coulomb divergence. In the case of the vector counterparts,
the leading contributions to $H_{\rm NRQCD}$ and $Q_{\rm NRQCD}$ are both
Coulomb divergent, which show the behavior $\propto 1/|\bm{q}|$ in the limit
$|\bm{q}|\to 0$. The expressions in (\ref{HQ-NRQCD-final}) reproduce the
IR behaviors of the full QCD counterparts in (\ref{HQ-final}).
\section{Results for the short distance coefficients\label{sec:results}}
In this section, we list our final results for the short distance coefficients
$P_n^{(j)}$, $a_n^{(j)}$, and $b_n^{(j)}$ for $j=0$ and 1 and for $n=0$, 1,
and 2. We have shown that the IR behaviors of $G_5$, $H_5$,
$Q_5$, $G$, $H$, and $Q$ in (\ref{G-final}) and (\ref{HQ-final}) are exactly
reproduced by the NRQCD counterparts $G_{5,{\rm NRQCD}}$, $H_{5,{\rm NRQCD}}$,
$Q_{5,{\rm NRQCD}}$, $G_{\rm NRQCD}$, $H_{\rm NRQCD}$, and $Q_{\rm NRQCD}$
in (\ref{G-NRQCD-final}) and (\ref{HQ-NRQCD-final}), respectively.
Therefore, all of the quantities $\Delta G_5$, $\Delta H_5$,
$\Delta Q_5$, $\Delta G$, $\Delta H$, and $\Delta Q$
in (\ref{DeltaAB}) are free of IR singularities.
Because the imaginary parts of the QCD amplitudes are the same as those of
the NRQCD counterparts, all of the quantities $\Delta G_5$, $\Delta H_5$,
$\Delta Q_5$, $\Delta G$, $\Delta H$, and $\Delta Q$ in (\ref{DeltaAB})
are real. Except for $\Delta G_5$ and $\Delta G$, which have logarithmic
UV divergences originated from the NRQCD factors $G_{5,{\rm NRQCD}}$ and
$G_{\rm NRQCD}$, all of the other quantities
($\Delta H_5$, $\Delta Q_5$, $\Delta H$, and $\Delta Q$) are finite in both
UV and IR regions. We renormalize the UV divergences of $\Delta G_5$ and
$\Delta G$ according to the $\overline{\rm MS}$ scheme. Our final results
for $\Delta G^{(1)}_{5,\overline{\rm MS}}$, $\Delta H^{(1)}_5$,
and $\Delta Q^{(1)}_5$ are
\begin{subequations}
\label{DG5-final}
\begin{eqnarray}
\Delta G^{(1)}_{5,\overline{\rm MS}} &=& \frac{\alpha_s C_F}{4\pi}
  \bigg\{- 4  + 2\bigg[3 + \frac{(m_b+m_c)^2}{p_{\bar{b}}\cdot p_c-m_bm_c} \bigg]
 \delta^2L_1(\delta,e_c)
-\frac{p_{\bar{b}}\cdot p_c}{2p^2}
\bigg(2K(\delta,e_c)
\nonumber\\
&&
+
\frac{L_2(\delta,e_c)}{2}
\log \frac{m_c^2}{m_b^2}
\bigg)
  +\bigg[\frac{p_{\bar{b}}\cdot p_c}{4p^2}
L_1(\delta,e_c)
  -1\bigg]\log \frac{\mu^4}{m_b^2m_c^2}
 -\frac{m_b^2-m_c^2}{8p^2} \log\frac{m_c^2}{m_b^2}
     \bigg\},
\nonumber\\ \\
\Delta H^{(1)}_5 &=& \frac{\alpha_sC_F}{2\pi}
\bigg[
    \frac{m_b-m_c}{p_{\bar{b}}\cdot p_c+m_bm_c}\,\delta^2L_1(\delta,e_c)
  + \frac{m_b+m_c}{8p^2}\log\frac{m_c^2}{m_b^2}\,
\bigg],
\\
\Delta Q^{(1)}_5 &=&\frac{\alpha_s C_F}{4\pi}
\bigg\{
  \bigg[
  \frac{4}{p_{\bar{b}}\cdot p_c-m_bm_c}
  - \frac{(m_b-m_c)^2}{2p^2(p_{\bar{b}}\cdot p_c+m_bm_c)}
  + \frac{1}{p^2}
  \bigg](m_b+m_c) \delta^2 L_1(\delta,e_c)  \nonumber \\
&&+ \bigg[3
  -\frac{(m_b+m_c)^2}{2p^2}\bigg]
    \frac{m_b-m_c}{4p^2}\log\frac{m_c^2}{m_b^2}
    - \frac{m_b+m_c}{p^2}
  \bigg\}.
\end{eqnarray}
\end{subequations}
The results for the vector part can be obtained by replacing $m_c$
in (\ref{DG5-final}) with $-m_c$ as
\begin{subequations}
\label{DG-final}
\begin{eqnarray}
\Delta G^{(1)}_{\overline{\rm MS}} &=&
\Delta G^{(1)}_{5,\overline{\rm MS}}\big|_{m_c\to-m_c},
\\
\Delta H^{(1)} &=&
\Delta H^{(1)}_{5}\big|_{m_c\to-m_c},
 \\
\Delta Q^{(1)} &=&
\Delta Q^{(1)}_{5}\big|_{m_c\to-m_c}.
\end{eqnarray}
\end{subequations}
Taking $m_b=m_c=m_{Q}$ in (\ref{DG-final}) for the vector current,
we recover the corresponding results of the electromagnetic current
for the spin-triplet $S$-wave $Q\bar{Q}$ pair of the same
flavor in \cite{Bodwin:2008vp}.

Now we are ready to obtain the short distance coefficients $P_n^{(j)}$,
$a_n^{(j)}$, and $b_n^{(j)}$ for $n=0$, 1, and 2 and for $j=0$ and 1.
The order-$\alpha_s^0$ short distance coefficients
can be found from the expansion formulas in (\ref{an0-bn0})
and (\ref{an0-bn0-2}) as
\begin{subequations}
\label{pab0-final}
\begin{eqnarray}
P_0^{(0)}&=&1,
\\
P_1^{(0)}&=&
- \frac{(m_b+m_c)^2}{8m_b^2m_c^2},
\\
P_2^{(0)}&=&
\frac{(m_b+m_c)^2}{128m_b^4m_c^4}(11m_b^2  -10m_b m_c+ 11m_c^2),
\\
a_0^{(0)}&=&1,
 \\
a_1^{(0)}&=&
- \frac{(m_b-m_c)^2}{8m_b^2m_c^2},
\\
a_2^{(0)}&=&
\frac{(m_b-m_c)^2}{128m_b^4m_c^4}(11m_b^2 +10m_b m_c+ 11m_c^2),
\\
b_0^{(0)}&=&-\frac{1}{2 m_b m_c},
 \\
b_1^{(0)}&=&
\frac{3(m_b^2 + m_c^2)}{16m_b^3 m_c^3},
\\
b_2^{(0)}&=&
-\frac{31 m_b^4 + 18 m_b^2 m_c^2 + 31 m_c^4}
 {256 m_b^5 m_c^5}.
\end{eqnarray}
\end{subequations}
It is straightforward to
obtain $a_n^{(0)}$ and $b_n^{(0)}$ for $n\ge 3$ in the same way.
The $S$- and $D$-wave components of the short distance coefficient for the
vector current can be obtained from (\ref{pab0-final})  as
$S_n^{(j)}=a_n^{(j)}+\tfrac{1}{3}b_n^{(j)}$ and $D_n^{(j)}=a_n^{(j)}$.
The short distance coefficients of order $\alpha_s^0$ are free of
scale dependence. The values for $P_0^{(0)}$ and $P_1^{(0)}$ agree with
the previous results in \cite{Braaten:1995ej,Hwang:1999fc,Onishchenko:2003ui}.
$S_0^{(0)}$ and $S_1^{(0)}$ agree with the previous results in
\cite{Hwang:1999fc}.

The order-$\alpha_s^1$ short distance coefficients
$P_n^{(1)}$ for $n=0$, 1, and 2 can be obtained by
substituting $[\Delta G_5^{(1)}]_{\rm \overline{MS}}$,
$\Delta H_5^{(1)}$, and $\Delta Q_5^{(1)}$ into
$\Delta G_5^{(1)}$,
$\Delta H_5^{(1)}$, and $\Delta Q_5^{(1)}$
in (\ref{an1-bn1}), respectively. The results are
\begin{subequations}
\label{pn1-final}
\begin{eqnarray}
P_0^{(1)}&=& \frac{\alpha_sC_F}{4\pi}
\bigg[-6 - \frac{3(m_b-m_c)}{2(m_b+m_c)} \log\frac{m_c^2}{m_b^2}\bigg],
\\
\big[P_1^{(1)}\big]_{\overline{\rm MS}}
&=&\frac{\alpha_sC_F}{4\pi m_b^2m_c^2}\bigg\{
\frac{-1}{144}\bigg[4(m_b^2+98m_bm_c+m_c^2) +
  3(7m_b^2+46m_bm_c+7m_c^2)
    \nonumber \\
&&\times
  \frac{m_b-m_c}{m_b+m_c} \log \frac{m_c^2}{m_b^2}
   \bigg]
+\frac{2}{3}(m_b+m_c)^2\log\frac{\mu^2}{m_bm_c}\bigg\},
\\
\big[P_2^{(1)}\big]_{\overline{\rm MS}}
&=&\frac{\alpha_sC_F}{4\pi m_b^4m_c^4}\bigg\{
\frac{-1}{57600}\bigg[4(4089m_b^4 -
  10364m_b^3m_c - 36586m_b^2m_c^2 - 10364m_bm_c^3
  \nonumber\\
&&  + 4089m_c^4)
  -15(177m_b^4 +
    2212m_b^3m_c + 4582m_b^2m_c^2 + 2212m_bm_c^3+177m_c^4)
  \nonumber\\
&& \times\frac{m_b-m_c}{m_b+m_c} \log \frac{m_c^2}{m_b^2}
    \bigg]
-
  \frac{(m_b+m_c)^2}{60}(21m_b^2+2m_bm_c+21m_c^2)
  \log\frac{\mu^2}{m_bm_c}\bigg\}.\phantom{xxxx}
\end{eqnarray}
\end{subequations}
Except for $P_0^{(1)}$, $P_n^{(1)}$ for $n\ge 1$ are dependent on
the NRQCD factorization scale $\mu$ that has been introduced
in the process of renormalization.
Our result for $P_0^{(1)}$ agrees with those in
\cite{Braaten:1995ej,Hwang:1999fc,Onishchenko:2003ui}.
The results for $\big[P_1^{(1)}\big]_{\overline{\rm MS}}$
and $\big[P_2^{(1)}\big]_{\overline{\rm MS}}$ are new.

The short distance coefficients $a_n^{(1)}$ and $b_n^{(1)}$ for $n=0$,
1, and 2 of the vector current at order $\alpha_s^1$ are obtained by
substituting $[\Delta G^{(1)}]_{\rm \overline{MS}}$
and $\Delta H^{(1)}$ into
$\Delta G^{(1)}$ and
$\Delta H^{(1)}$ in (\ref{an1-bn1-2}), respectively.
The results for $a_n^{(1)}$ are
\begin{subequations}
\label{a1-final}
\begin{eqnarray}
a_0^{(1)}&=&\frac{\alpha_sC_F}{4\pi}
\bigg[ -8 -\frac{3(m_b-m_c)}{2(m_b+m_c)} \log\frac{m_c^2}{m_b^2}\bigg],
\\
{[}a_1^{(1)}]_{\overline{\rm MS}}
&=&\frac{\alpha_sC_F}{4\pi m_b^2m_c^2}
\bigg\{ \frac{1}{144}
\bigg[32(m_b^2 - m_bm_c + m_c^2) - 3
   (7m_b^2 + 10m_bm_c + 7m_c^2)   \nonumber \\
 &&\times
 \frac{m_b-m_c}{m_b+m_c} \log \frac{m_c^2}{m_b^2}
 \bigg]
+ \frac{2}{3}(m_b+m_c)^2\log\frac{\mu^2}{m_bm_c}\bigg\}
,
\\
{[}a_2^{(1)}]_{\overline{\rm MS}}
&=&\frac{\alpha_sC_F}{4\pi m_b^4m_c^4}\bigg\{
  \frac{-1}{19200} \bigg[16(547m_b^4 + 128 m_b^3m_c +
  122m_b^2m_c^2 + 128m_bm_c^3+547m_c^4)
\nonumber\\
&& - 15 (59m_b^4 + 164m_b^3m_c + 194m_b^2m_c^2 +
  164m_bm_c^3 + 59m_c^4)
\frac{m_b-m_c}{m_b+m_c} \log\frac{m_c^2}{m_b^2}
 \bigg]
\nonumber\\&&
 - \frac{1}{20}(m_b+m_c)^2(7m_b^2 - 6m_bm_c + 7m_c^2)
 \log\frac{\mu^2}{m_bm_c}
\bigg\}.
\end{eqnarray}
\end{subequations}
And the short distance coefficients $b_n^{(1)}$ are
\begin{subequations}
\label{ab1-final}
\begin{eqnarray}
b_0^{(1)}&=&\frac{\alpha_sC_F}{4\pi m_bm_c}\bigg[
2 +\frac{m_b-m_c}{4(m_b+m_c)} \log
  \frac{m_c^2}{m_b^2}\bigg],
 \\
{[}b_1^{(1)}]_{\overline{\rm MS}}
&=&\frac{\alpha_sC_F}{4\pi m_b^3m_c^3}\bigg\{
\frac{-1}{288}\bigg[8(19m_b^2 - 10 m_bm_c +
  19m_c^2) - 21
  (m_b^2 + 4m_bm_c + m_c^2)
\nonumber\\
&&\times
 \frac{m_b-m_c}{m_b+m_c}\log \frac{m_c^2}{m_b^2}\bigg]
-\frac{1}{3}(m_b+m_c)^2\log \frac{\mu^2}{m_bm_c}
\bigg\},
\\
{[}b_2^{(1)}]_{\overline{\rm MS}}
&=&\frac{\alpha_sC_F}{4\pi m_b^5m_c^5}\bigg\{
\frac{1}{115200} \bigg[8(5647m_b^4 - 972m_b^3m_c +
  922m_b^2m_c^2 - 972m_bm_c^3 + 5647m_c^4)  \nonumber \\
& & -15(527m_b^4 + 1912m_b^3m_c + 2082m_b^2m_c^2
    + 1912m_bm_c^3 + 527m_c^4)
\frac{m_b-m_c}{m_b+m_c}\log \frac{m_c^2}{m_b^2}\bigg]
\nonumber \\
  &&+
  \frac{1}{120}(m_b+m_c)^2(31m_b^2 - 8m_bm_c +
  31m_c^2)\log \frac{\mu^2}{m_bm_c}\bigg\}.
\end{eqnarray}
\end{subequations}
As in the case of $a_n^{(0)}$ and $b_n^{(0)}$, it is straightforward to
obtain $a_n^{(1)}$ and $b_n^{(1)}$ for $n\ge 3$.
Except for $a_0^{(1)}$ and $b_0^{(1)}$,
$a_n^{(1)}$ and $b_n^{(1)}$ for $n\ge 1$ are dependent on
the NRQCD factorization scale $\mu$.
Our result for $S_0^{(1)}=a_0^{(1)}+\tfrac{1}{3}b_0^{(1)}$
agrees with that in \cite{Hwang:1999fc}.
The results for
$\big[a_1^{(1)}\big]_{\overline{\rm MS}}$,
$\big[b_1^{(1)}\big]_{\overline{\rm MS}}$,
$\big[a_2^{(1)}\big]_{\overline{\rm MS}}$, and
$\big[b_2^{(1)}\big]_{\overline{\rm MS}}$
are new.

Let us make a rough estimate of the effect of relativistic corrections at
order $\alpha_s$. It is convenient to define the ratio of
the NRQCD matrix element of relative order-$\bm{q}^{2n}$ to the leading-order
matrix element:
\begin{equation}
\langle \bm{q}^{2n}\rangle_{B_c}\equiv
\frac{\langle 0|{\cal O}_n|B_c\rangle}
     {\langle 0|{\cal O}_0|B_c\rangle},
     \,\,\,\,\,\,\,
\langle \bm{q}^{2n}\rangle_{B_c^*}\equiv
\frac{\langle 0|{\cal O}^i_{An}|B_c^*\rangle}
     {\langle 0|{\cal O}^i_{A0}|B_c^*\rangle},
\end{equation}
where ${\cal O}_n$ and ${\cal O}^i_{An}$ are defined in (\ref{operators}), and
we have used the property that $\langle \bm{q}^{2n}\rangle_{B_c^*}$
is independent of $i$.
The ratios $\langle \bm{q}^{2n}\rangle_{B_c,B_c^*}$
are normalized to be consistent with that, $\langle \bm{q}^{2n}\rangle_{H}$,
for the $Q\bar{Q}$ quarkonium $H$ considered in
\cite{Bodwin:2006dn,Bodwin:2007fz,Bodwin:2008vp}.
In \cite{Bodwin:2006dn}, it was shown that the $\overline{\rm MS}$ value
for $\langle \bm{q}^{2n}\rangle_{H}$
satisfies a generalized Gremm-Kapustin relation \cite{Gremm:1997dq}:
\begin{equation}
{[}\langle \bm{q}^{2n}\rangle]_{\overline{\rm MS}}
={[}\langle \bm{q}^2\rangle]_{\overline{\rm MS}}^n.
\end{equation}
We assume that this relation is still valid in the case of the $\bar{b}c$
meson. Unfortunately, unlike the case of the $S$-wave bound states of
the $Q\bar{Q}$ pairs, the leptonic decay rates for $B_c$ and $B_c^*$
have not been measured accurately so that one cannot determine
$\langle \bm{q}^{2n}\rangle_{B_c,B_c^*}$ with empirical data. Instead, by taking
upper and lower bounds of the ratios $\langle \bm{q}^{2n}\rangle_{B_c,B_c^*}$
as $\langle \bm{q}^{2}\rangle_{J/\psi}=0.441\,\textrm{GeV}^2$ \cite{Bodwin:2007fz}
and  $\langle \bm{q}^{2}\rangle_{\Upsilon(1S)}=-0.193\,\textrm{GeV}^2$ \cite{Chung:2010vz},
respectively, we make a rough estimate of the sums of products of $S$-wave
short distance coefficients and operator matrix elements.
Taking the central values for the variables $m_b=4.6\,{\rm GeV}$,
$m_c=1.5\,{\rm GeV}$, $\mu=(m_b+m_c)/2=3.05\,{\rm GeV}$, and
$\langle{\bm q}^2\rangle_{B_c,B_c^*}=(\langle \bm{q}^{2}\rangle_{J/\psi}+%
\langle \bm{q}^{2}\rangle_{\Upsilon(1S)})/2=0.124\,{\rm GeV}^2$,
we find that
\begin{subequations}
\label{num-coefficient}%
\begin{eqnarray}
\sum_{n=0}^0\left[P_n^{(1)}\right]_{\overline{\rm MS}}
\,
[\langle\bm{q}^{2}\rangle_{B_c}]^{n}_{\overline{\rm MS}}
&=&
- \frac{\alpha_s C_F}{4 \pi}\times
4.292
,
\\
\sum_{n=0}^1\left[P_n^{(1)}\right]_{\overline{\rm MS}}
\,
[\langle\bm{q}^{2}\rangle_{B_c}]^{n}_{\overline{\rm MS}}
&=&
- \frac{\alpha_s C_F}{4 \pi}\times
4.293^{+0.185}_{-0.330}
,
\\
\sum_{n=0}^2\left[P_n^{(1)}\right]_{\overline{\rm MS}}
\,
[\langle\bm{q}^{2}\rangle_{B_c}]^{n}_{\overline{\rm MS}}
&=&
- \frac{\alpha_s C_F}{4 \pi}\times
4.294^{+0.152}_{-0.306}
,
\\
\label{coeff-infty}
\sum_{n=0}^\infty\left[P_n^{(1)}\right]_{\overline{\rm MS}}
\,
[\langle\bm{q}^{2}\rangle_{B_c}]^{n}_{\overline{\rm MS}}
&=&
- \frac{\alpha_s C_F}{4 \pi}\times
4.294^{+0.152}_{-0.308}
,
\\
\sum_{n=0}^0\left[S_n^{(1)}\right]_{\overline{\rm MS}}
\,
[\langle\bm{q}^{2}\rangle_{B_c^*}]^{n}_{\overline{\rm MS}}
&=&
- \frac{\alpha_s C_F}{4 \pi}\times
6.209
,
\\
\sum_{n=0}^1\left[S_n^{(1)}\right]_{\overline{\rm MS}}
\,
[\langle\bm{q}^{2}\rangle_{B_c^*}]^{n}_{\overline{\rm MS}}
&=&
- \frac{\alpha_s C_F}{4 \pi}\times
6.168^{+0.214}_{-0.289}
,
\\
\sum_{n=0}^2\left[S_n^{(1)}\right]_{\overline{\rm MS}}
\,
[\langle\bm{q}^{2}\rangle_{B_c^*}]^{n}_{\overline{\rm MS}}
&=&
- \frac{\alpha_s C_F}{4 \pi}\times
6.170^{+0.213}_{-0.247}
,
\\
\sum_{n=0}^\infty\left[S_n^{(1)}\right]_{\overline{\rm MS}}
\,
[\langle\bm{q}^{2}\rangle_{B_c^*}]^{n}_{\overline{\rm MS}}
&=&
- \frac{\alpha_s C_F}{4 \pi}\times
6.170^{+0.213}_{-0.252} ,
\end{eqnarray}
\end{subequations}
where the uncertainties are estimated by varying the values
$m_c\le \mu\le m_b$ and
$-0.193\,\textrm{GeV}^2$ $\le\langle\bm{q}^{2}\rangle_{B_c}\le%
0.441\,\textrm{GeV}^2$. The estimates in (\ref{num-coefficient})
show that the series expansions for the relativistic corrections to the
axial vector and vector currents at order $\alpha_s$ converge rapidly
in spite of the large uncertainties in the ratio
$\langle\bm{q}^{2}\rangle_{B_c}$. In addition, the order-one contributions
of the form $(m_b-m_c)/(m_b+m_c) \log (m_c^2/m_b^2)$ do not deteriorate the
convergence of the short distance coefficients for the axial vector and
vector currents. In \cite{Onishchenko:2003ui}, the authors obtained the
short distance coefficient $P_0^{(2)}$ of order-$\alpha_s^2 v^0$ as
$-24.4\times[\alpha_s(m_b)/\pi]^2$ with the input parameters
$m_b = \mu = 4.8\,$GeV and $m_c =1.65\,$GeV. If we use these
values for $m_b$, $m_c$, $\mu$ and set
$\langle{\bm q}^2\rangle_{B_c}=0.124\,{\rm GeV}^2$, then
the coefficients (\ref{num-coefficient}) vary by about $\lesssim 3\,\%$.
As an example, (\ref{coeff-infty}) becomes
\begin{equation}
\sum_{n=0}^\infty\left[P_n^{(1)}\right]_{\overline{\rm MS}}
\,
[\langle\bm{q}^{2}\rangle_{B_c}]^{n}_{\overline{\rm MS}}
=
- \frac{\alpha_s C_F}{4 \pi}\times
4.399.
\end{equation}
\section{Summary\label{sec:summary}}
We have computed the short distance coefficients for the NRQCD factorization
formulas of the meson-to-vacuum matrix elements for the axial vector and
vector parts of the charged weak current in the $S$-wave spin-singlet and
-triplet $\bar{b}c$ mesons, respectively. The computation was carried out
at order $\alpha_s^0$ and $\alpha_s^1$ including relativistic corrections
of all of the $\bar{b}c$ NRQCD operators that contain any number of ordinary
derivatives without gauge fields. In the Coulomb gauge, gauge field
contributions first appear from relative order $v_Q^4$. The numerical value
for the correction of order $\alpha_sv_Q^4$ is tiny ($< $0.1\% of the
leading order contribution). We have reproduced all available short distance
coefficients of order $\alpha_s$ or less and our results of order $\alpha_s^1$
with relativistic corrections are new. By taking the limit $m_b=m_c=m_Q$,
we have reproduced the results for the order-$\alpha_s$ corrections to the
quarkonium electromagnetic current for the spin-triplet $S$-wave
$Q\bar{Q}$ pair with the same flavor.
Although we have listed explicitly the first few short distance coefficients
of order-$\alpha_s$ for the relativistic corrections, it is straightforward
to obtain the terms of higher orders in $\bm{q}^{2n}$. The results reveal that,
in spite of the large uncertainties in the ratios
$\langle\bm{q}^{2}\rangle_{B_c,B_c^*}$, the relativistic corrections to the
leptonic decay rate of the $B_c(Bc^*)$ meson at order $\alpha_s^1$ or less converge
rapidly, which shows a strong contrast to the uncomfortably large corrections
of order $\alpha_s^2|\bm{q}|^0$. The short distance coefficients listed in this
paper can be employed to compute the resummation of relativistic corrections
to the phenomenological measurables that involve $B_c$ and $B_c^*$
production and decay.
\begin{acknowledgments}
S.K. was supported by the National Research Foundation of Korea grant funded
by the Korea government (MEST) No. 2010-0022219.
The work of J.L. and W.S. was
supported by Basic Science Research Program through
the NRF of Korea funded by the MEST under contracts
2010-0015682 (J.L.) and 2010-0027811 (W.S.).
J.L. and W.S. were also supported in part by a Korea University fund.
\end{acknowledgments}
\appendix
\section{Tensor-integral reduction\label{appendix:tensor}}
In this appendix, we describe the tensor-integral reduction that we use to
simplify (\ref{QCD-lambda}) to obtain (\ref{l5mu-lmu}).

Tensor integrals of rank 1 and 2 that depend on $p$ or on
both $p$ and $q$ can be expressed in terms of scalar integrals as follows:
\begin{eqnarray}
\label{kmu-p-only}%
\int_k k^\mu f(k,p)&=&
\frac{p^\mu}{p^2}\int_k p\cdot k f(k,p),
\\
\int_k k^\mu k^\nu f(k,p)&=&
g^{\mu\nu}\int_k d_1(k,p) f(k,p)
+p^\mu p^\nu \int_k d_2(k,p)f(k,p),
\\
\int_k k^\mu f(k,p,q)&=&
p^\mu\int_k d_3(k,p,q) f(k,p,q)
+
q^\mu \int_k d_4(k,p,q) f(k,p,q),
\\
\int_k k^\mu k^\nu f(k,p,q)&=&
g^{\mu\nu}\int_k d_5(k,p,q)f(k,p,q)
+
p^\mu p^\nu\int_k d_6(k,p,q)f(k,p,q)
\nonumber\\
&&
+
q^\mu q^\nu\int_k d_7(k,p,q)f(k,p,q)
\nonumber\\
&&
+
(p^\mu q^\nu + p^\nu q^\mu)\int_k d_8(k,p,q)f(k,p,q),
\end{eqnarray}
where $k$ is the loop momentum, the symbol $\int_k$ is defined in
(\ref{intk-di}), and $f$ is an arbitrary scalar function of the argument
four-vectors. The functions $d_i$'s are defined by
\begin{eqnarray}
d_1(k,p)&=&\frac{1}{d-1}\left[k^2-\frac{(k\cdot p)^2}{p^2}\right],
\\
d_2(k,p)&=&\frac{1}{(d-1)p^2}\left[-k^2+d\frac{(k\cdot p)^2}{p^2}\right],
\\
d_3(k,p,q)
&=&
\frac{k\cdot q\,p\cdot q-q^2\,k\cdot p}{(p\cdot q)^2-p^2q^2},
\\
d_4(k,p,q)
&=&d_3(k,q,p),
\\
d_5(k,p,q)
&=&\frac{1}{d-2}
 \left[ k^2 + \frac{p^2 (q \cdot k)^2 + q^2 (p \cdot k)^2
                   - 2\,p \cdot q\,p \cdot k\,q \cdot k
   }{(p\cdot q)^2 - p^2 q^2}
  \right],
\\
d_6(k,p,q)
&=&\frac{1}{(d-2)\left[(p\cdot q)^2-p^2 q^2\right]^2}
\bigg\{\big[(k\cdot p)^2 (d-1)-p^2 k^2\big]q^4
  +(d-2)(p\cdot q)^2 (k\cdot q)^2
   \nonumber\\
   &&
   +\big[k^2 (p\cdot q)^2-2(d-1) (k\cdot p) (k\cdot q)
   (p\cdot q)
   +p^2 (k\cdot q)^2\big] q^2
   \bigg\},
\\
d_7(k,p,q)
&=&d_6(k,q,p),
\\
d_8(k,p,q)
&=&\frac{1}{(d-2)\left[(p\cdot q)^2-p^2
   q^2\right]^2 }\bigg\{\left[q^2 (k\cdot p)^2-k^2 (p\cdot q)^2+p^2
   (k\cdot q)^2+p^2 q^2 k^2\right] p\cdot q\nonumber\\
   &-&2 p^2
   q^2 \,k\cdot p\,k\cdot q+d\left[p\cdot q\,k\cdot p-p^2 \,k\cdot q
                             \right]
   \left[p\cdot q\,k\cdot q-q^2 \,k\cdot p
   \right]
\bigg\}.
\end{eqnarray}
\section{Scalar integrals for the vertex corrections\label{appendix:scalarint}}
In this appendix, we list the definitions and the values for the scalar
integrals $J_i$ that appear in the vertex corrections in (\ref{l5mu-lmu}).
The scalar integrals $J_i$'s are defined by
\begin{equation}
J_i=\int_k \frac{N_i}{D_0 D_1 D_2},
\end{equation}
where $k$ is the loop momentum, the symbol $\int_k$ and the denominator
factors $D_i$'s are defined in (\ref{intk-di}), and the numerators
$N_i$'s of the integrand are defined by
\begin{eqnarray}
N_1 &=& k^2,
\\
N_2 &=& 1,
\\
N_3 &=& 2 k \cdot q,
\\
N_4 &=& \frac{1}{d-2}
 \Bigg[ k^2 + \frac{p^2 (q \cdot k)^2 + q^2 (p \cdot k)^2
                   - 2(p \cdot q) (p \cdot k) (q \cdot k)
   }{(p\cdot q)^2 - p^2 q^2}
  \Bigg], \\
N_5 &=& 2 k\cdot p,
\end{eqnarray}
\begin{eqnarray}
N_6 &=& \frac{m_b+m_c}{d-2} \;\frac{q^2}{(p \cdot q)^2 -
  p^2 q^2}
  \Bigg\{
  k^2 p^2 - (p \cdot k)^2 +
  \frac{(d - 1)[p^2 (q \cdot k) - (p \cdot q) (p
  \cdot k)]^2}{(p\cdot q)^2 - p^2 q^2}  \nonumber \\
&+& \frac{m_c - m_b}{m_c +  m_b} \Bigg[- k^2 (p \cdot q) + (p
  \cdot k)(q \cdot k) + \frac{(d - 1)[q^2 (p
  \cdot k) - (p \cdot q)(q \cdot k)]}
  {(p \cdot q)^2 - p^2 q^2}   \nonumber \\
&&\qquad\qquad\quad
\times[p^2 (q \cdot k) - (p \cdot q)(p \cdot k) ]\Bigg]
\Bigg\},\\
N_7 &=&  \frac{m_b+m_c}{d-2} \;\frac{p^2 q^2}{(p\cdot q)^2 - p^2 q^2}
  \Bigg\{ - k^2 (p \cdot q) + (p \cdot k)(q\cdot k)
\nonumber \\
&+& \frac{(d - 1)[q^2 (p \cdot k) -(p \cdot q)(q \cdot k)]
                 [p^2 (q \cdot k) - (p \cdot q)(p \cdot k)]}
         {(p \cdot q)^2 - p^2 q^2}
\nonumber \\
&+& \frac{m_c - m_b}{m_c + m_b}\Bigg[k^2 q^2 - (q \cdot k)^2
 + \frac{(d-1)[q^2 (p \cdot k) - (p \cdot q) (q\cdot k)]^2}
        {(p \cdot q)^2 - p^2 q^2}
  \Bigg]
  \Bigg\},
\end{eqnarray}
where $d=4-2\epsilon$ is the number of space-time dimensions.
The external momenta $p$ and $q$ are
defined by $p=\tfrac{1}{2}(p_c+p_{\bar{b}})$
and $q=\tfrac{1}{2}(p_c-p_{\bar{b}})$,
where $p_{\bar{b}}$ and $p_c$ are the momenta for the $\bar{b}$
and $c$, respectively, which are on their mass shells:
$p_{\bar{b}}^2=m_b^2$ and $p_{c}^2=m_c^2$.

The values for the scalar integrals $J_i$'s are
\begin{eqnarray}
J_1 &=&
\frac{i}{(4\pi)^2} \Bigg[\frac{1}{\epsilon_{\rm UV}}+2+2 i\pi \delta - 2\delta^2
  L_1(\delta,e_c)
  + e_c\log \frac{4\pi\mu^2 e^{-\gamma_{\rm E}}}{m_c^2}
\nonumber\\
&&  + (1-e_c)\log \frac{4\pi\mu^2 e^{-\gamma_{\rm E}}}{m_b^2}
\Bigg],
\\
J_2&=&\frac{i}{(4\pi)^2}\;\frac{1}{8p^2}
  \Bigg\{
  \bigg(\frac{1}{\epsilon_{\rm IR}}
  +\frac{1}{2}\log \frac{(4\pi\mu^2 e^{-\gamma_{\rm E}})^2}{m_b^2 m_c^2}
  \bigg)
  \bigg[\frac{i\pi}{\delta} - L_1(\delta,e_c)
  \bigg]
   + 2 K(\delta,e_c) - \frac{\pi^2}{\delta} \nonumber \\
&&+ \frac{i\pi}{2\delta}\log \bigg[\frac{p^4 m_b^2 m_c^2}
                      {[(p_{\bar{b}} \cdot  p_c)^2-m_b^2 m_c^2]^2}\bigg]
   + \frac{1}{2}L_2(\delta,e_c)\log \frac{m_c^2}{m_b^2}
  \Bigg\},
\\
J_3&=& - \frac{i}{(4\pi)^2}\Bigg[2i\pi \delta - 2 \delta^2 L_1(\delta,e_c)
+  \frac{m_b^2-m_c^2}{8p^2} \log \frac{m_c^2}{m_b^2}\Bigg],
\\
J_4 &=& \frac{1}{4}\Bigg[\frac{i}{(4\pi)^2}+J_1\Bigg],
\\
J_5&=& \frac{i}{2(4\pi)^2} \,
    \log\frac{m_c^2}{m_b^2},
\\
J_6
&=& \frac{i}{(4\pi)^2}
 \frac{(m_b+m_c)q^2}{p_{\bar{b}} \cdot p_c - m_b m_c}
 \!
  \Bigg[\delta^2 L_1(\delta,e_c) -i\pi \delta
  + \frac{1}{4}\bigg(\frac{m_b-m_c}{m_b+m_c} -
  \frac{m_b^2 - m_c^2}{4p^2}\bigg)
  \log \frac{m_c^2}{m_b^2}\Bigg],
\phantom{xxxxx}
\end{eqnarray}
\begin{eqnarray}
J_7 &=&\frac{i}{4(4\pi)^2}
\frac{(m_b+m_c)q^2}{p_{\bar{b}} \cdot p_c - m_b  m_c}
  \Bigg\{
  \bigg[m_b^2-m_c^2+2(p_{\bar{b}} \cdot p_c - m_bm_c)
   \frac{m_b-m_c}{m_b+m_c}
  \bigg]
  \big[\delta^2 L_1(\delta,e_c)-i\pi \delta\big]
  \nonumber \\
&&+\bigg[\frac{m_b^2-m_c^2}{4}
  \bigg(\frac{m_b-m_c}{m_b+m_c} -\frac{m_b^2-m_c^2}{4p^2}\bigg)
-  \frac{p_{\bar{b}} \cdot p_c - m_b m_c}{2}
  \bigg(1-\frac{m_b^2-m_c^2}{4p^2}\bigg)
  \bigg]
  \nonumber\\
  &&
  \times\log\frac{m_c^2}{m_b^2}\Bigg\},
\end{eqnarray}
where the variables $\delta$ and $e_c$ are defined in (\ref{def-ab})
and the functions $L_1(\delta,e_c)$, $L_2(\delta,e_c)$, and $K(\delta,e_c)$
are defined in (\ref{def-LK}).

\section{Integrals for NRQCD vertex corrections\label{appendix:NRscalarint}}
In this appendix, we list elementary loop integrals that are useful
in computing the NRQCD corrections considered in section~\ref{sec:NRQCD}.
We also list the values for the integrals defined in
(\ref{S-integrals}). We follow the strategy of evaluating the integrals
given in \cite{Bodwin:2008vp}.
\subsection{Elementary scalar integrals}
In dimensional regularization, scaleless power-divergent integrals vanish:
\begin{equation}
\label{scaleless0}%
\int_{\bm{k}} \frac{1}{|\bm{k}|^{n}}
=0
\end{equation}
for $n\neq 3$. The only nonvanishing scaleless integral is
\begin{equation}
\label{n0}%
n_0\equiv
\int_{\bm{k}} \frac{1}{|\bm{k}|^3}
=\frac{1}{4\pi^2}\left(
 \frac{1}{\epsilon_{\textrm{UV}}}
-\frac{1}{\epsilon_{\textrm{IR}}}\right),
\end{equation}
which diverges logarithmically. Except for the integral (\ref{n0}),
nonvanishing integrals are depending on $|\bm{q}|$, which are
\begin{subequations}
\label{n123}
\begin{eqnarray}
\label{n1}%
n_1&\equiv&
\int_{\bm{k}}
\frac{1}{\bm{k}^2+2\bm{k}\cdot \bm{q}-i\varepsilon}
=\frac{i}{4\pi}|\bm{q}|,
\\
\label{n2}%
n_2&\equiv&
\int_{\bm{k}}
\frac{1}{\bm{k}^2(\bm{k}^2+2\bm{k}\cdot \bm{q}-i\varepsilon)}
\nonumber\\
&=&-\frac{i}{16\pi|\bm{q}|} \left(
\frac{1}{\epsilon_{\textrm{IR}}}
+\log
\frac{\pi \mu^2 e^{-\gamma_{_{\!\textrm{E}}}} }{\bm{q}^2}
+i\pi
                  \right),
\\
\label{n3}%
n_3&\equiv&
\int_{\bm{k}}
\frac{\bm{k}^2}{\bm{k}^2+2\bm{k}\cdot \bm{q}-i\varepsilon}
=\frac{i}{2\pi}|\bm{q}|^3.
\end{eqnarray}
\end{subequations}

In projecting out the $S$-wave contribution from a scalar integral
that depends on $\bm{q}$, we have to take the average over the angle
of $\bm{q}$. The following formulas are useful in that step:
\begin{eqnarray}
\label{angularav1}%
\int_{\bm{k}}
\frac{f(\bm{k}^2)}{E_Q \pm\bm{q}\cdot\hat{\bm{k}}}
&=&
\frac{1}{2|\bm{q}|} \log \left(\frac{E_Q+|\bm{q}|}{E_Q-|\bm{q}|} \right)
\int_{\bm{k}}f(\bm{k}^2),
\\
\label{angularav2}%
\int_{\bm{k}}
\frac{f(\bm{k}^2)}{(E_b - \bm{q}\cdot\hat{\bm{k}})
                   (E_c + \bm{q}\cdot\hat{\bm{k}})}
&=&
\frac{1}{E_b+E_c}
\int_{\bm{k}} f(\bm{k}^2) \left(
\frac{1}{E_b+ \bm{q}\cdot\hat{\bm{k}}}
+
\frac{1}{E_c- \bm{q}\cdot\hat{\bm{k}}}
\right)
\nonumber\\
&=&
\frac{1}{2(E_b+E_c)|\bm{q}|}
\log \left[\frac{(E_b+|\bm{q}|)(E_c+|\bm{q}|)}
                {(E_b-|\bm{q}|)(E_c-|\bm{q}|)} \right]
\int_{\bm{k}}f(\bm{k}^2),\phantom{xxxx}
\end{eqnarray}
where $Q=b$ or $c$,
$\hat{\bm{k}}=\bm{k}/|\bm{k}|$, and
$f(\bm{k}^2)$  is any function of $\bm{k}^2$.

In the following sections, we express the integrals $S_1$, $S_2$, $S_3^\mu$,
and $S_4^{\mu\nu}$ defined in (\ref{S-integrals}) in linear combinations
of $n_0$, $n_1$, $n_2$, and $n_3$ in (\ref{n0}) and (\ref{n123}). We also
find the covariant forms of the integrals $S_1$, $S_2$, $S_3^\mu$, and
$S_4^{\mu\nu}$.
\subsection{$\bm{S_1}$}
The $k^0$ integral of $S_1$ defined in (\ref{S-integrals}) is
the sum of two contributions: $S_1=S_{1c}+S_{1\bar{b}}$, where
$S_{1c}$ and $S_{1\bar{b}}$ are the contributions from the
poles of the charm quark and antibottom quark, respectively.

The contribution from the $c$ pole is
\begin{equation}
S_{1c}
= - \frac{i}{4(E_b+E_c)} \mathcal{N}\!\!\!\!\!\!\!\int_{\bm{k}}
\frac{1}{\Delta_c(\Delta_c+E_c)}.
\end{equation}
We find that all of the factors in the denominator of the integrands
are of order $m_Q$ as $\bm{q}\to 0$ and $\bm{k}\to 0$. Therefore,
the expansion of $\Delta_Q$ in powers of $({\bm k}+{\bm q})^2/m_Q^2$
gives only scaleless, power-divergent integrals, which vanish, so that
\begin{equation}
\label{s1c}
S_{1c}=0.
\end{equation}

The contribution from the $\bar{b}$ pole is
\begin{equation}
\label{s1b}
S_{1\bar{b}}
=\frac{i}{4(E_b+E_c)} \mathcal{N}\!\!\!\!\!\!\!\int_{\bm{k}}
      \frac{1+(E_b/\Delta_b)}
  {\bm{k}^2+2\bm{k} \cdot \bm{q} - i \varepsilon}.
\end{equation}
Expanding $\Delta_Q$ in powers of $({\bm k}+{\bm q})^2/m_Q^2$, we obtain
\begin{equation}
\label{s1b1}
S_{1\bar{b}}
= \frac{i}{2(E_b+E_c)}n_1=-\frac{|{\bm q}|}{8\pi(E_b+E_c)},
\end{equation}
By adding the two contributions (\ref{s1c}) and (\ref{s1b1}), we obtain
\begin{equation}
\label{s1-final}
S_1
= \frac{i}{(4\pi)^2} 2\pi i \delta,
\end{equation}
where $\delta$ is defined in (\ref{def-ab}).
\subsection{$\bm{S_2}$}
The $k^0$ integral of $S_2$ is the sum of three contributions:
$S_2=S_{2g}+S_{2c}+S_{2\bar{b}}$, where
$S_{2g}$, $S_{2c}$, and $S_{2\bar{b}}$ are the contributions from the
poles of the gluon, charm quark, and antibottom quark, respectively.

The gluon pole contribution is
\begin{eqnarray}
\label{s2g}
S_{2g}
&=& \frac{i}{8} \mathcal{N}\!\!\!\!\!\!\!
\int_{\bm{k}} \frac{1}{|\bm{k}|^3(E_b-\bm{q}\cdot
  \hat{\bm{k}})(E_c + \bm{q} \cdot \hat{\bm{k}})}\nonumber\\
&=&\frac{i}{16(E_b+E_c)|\bm{q}|}
\,
n_0 \log \left[\frac{(E_b+|\bm{q}|)(E_c+|\bm{q}|)}
{(E_b-|\bm{q}|)(E_c-|\bm{q}|)} \right],
\end{eqnarray}
which is proportional to the scaleless logarithmically divergent integral
$n_0$.

The contribution from the $c$ pole is
\begin{equation}
\label{s2cexpnad}
S_{2c}
=
\frac{i}{4(E_b+E_c)} \mathcal{N}\!\!\!\!\!\!\!\int_{\bm{k}}
\frac{1}{\Delta_c (\Delta_c+E_c)
[\bm{k}^2 -(\Delta_c+E_c)^2 -i \varepsilon]}.
\end{equation}
We find that the factors $\Delta_c$ and $\Delta_c+E_c$ are of order
$m_Q$ as $\bm{q}\to 0$ and $\bm{k}\to 0$. Therefore, the expansion of the
factors $1/\Delta_c$ and $1/(\Delta_c+E_c)$ are trivial and the expansion
gives only scaleless power-divergent integrals. The expansion of the last
factor $1/[\bm{k}^2 -(\Delta_c+E_c)^2 -i \varepsilon]$ can be done in
powers of $\bm{k}^2/(\Delta_c+E_c)^2$ and then the factor
$1/(\Delta_c+E_c)^{2n}$ is expanded in powers of $({\bm k}+{\bm q})^2/m_Q^2$.
We find that all of the contributions are scaleless, power-divergent integrals
so that
\begin{equation}
\label{s2c}
S_{2c}=0.
\end{equation}

The contribution from the $\bar{b}$ is
\begin{equation}
\label{s2bexpand}
S_{2\bar{b}}
=
- \frac{i}{4(E_b+E_c)} \mathcal{N}\!\!\!\!\!\!\!\int_{\bm{k}}
\frac{1+(E_b/\Delta_b)}
     {[\bm{k}^2 - (\Delta_b-E_b)^2 -i\varepsilon]
      [\bm{k}^2+2\bm{k} \cdot \bm{q} - i \varepsilon]}.
\end{equation}
The expansion of the integrand for $S_{2\bar{b}}$ is similar to that
used in deriving (67) of \cite{Bodwin:2008vp}.
Following that method, we find that
\begin{equation}
\label{s2b}
S_{2\bar{b}}
=
-\frac{i}{2(E_b+E_c)}n_2=-\frac{1}{32\pi (E_b+E_c)|\bm{q}|}\left(
\frac{1}{\epsilon_{\textrm{IR}}}
+\log
\frac{\pi \mu^2 e^{- \gamma_{_{\!\textrm{E}}}}  }
{\bm{q}^2}
+i\pi
\right).
\end{equation}

The sum of the three contributions in (\ref{s2g}),
(\ref{s2c}), and (\ref{s2b}) is
\begin{equation}
\label{s2-final}
S_{2}
=\frac{i}{(4\pi)^2} \frac{1}{8p^2}
\bigg[\bigg(\frac{1}{\epsilon_{\rm UV}}
          - \frac{1}{\epsilon_{\rm IR}}\bigg) L_1(\delta,e_c)
     -\frac{\pi^2}{\delta} + \frac{i\pi}{\delta}
     \bigg(\frac{1}{\epsilon_{\rm IR}} + \log
  \frac{\pi \mu^2 e^{-\gamma_{\rm E}}}{{\bm{q}\;}^2}\bigg)
\bigg],
\end{equation}
where $\delta$ and $e_c$ are defined in (\ref{def-ab})
and the function $L_1(\delta,e_c)$ is defined in (\ref{def-LK}).
\subsection{$\bm{S_3^\mu}$}
The integral $S_3^\mu$ is the sum of three contributions:
$S_3^\mu=S_{3g}^\mu+S_{3c}^\mu+S_{3\bar{b}}^\mu$, where
$S_{3g}^\mu$, $S_{3c}^\mu$, and $S_{3\bar{b}}^\mu$ are the contributions
from the poles of the gluon, charm quark, and antibottom quark,
respectively.

Following the same way that has been used to evaluate the $k^0$
integrals of $S_1$ and $S_2$, we carry out the $k^0$ integrals
for $S_{3g}^\mu$, $S_{3c}^\mu$, and $S_{3\bar{b}}^\mu$ by contour
integration. The sum of the three contributions is
\begin{eqnarray}
S_3^0
&=&- \frac{i}{8}\mathcal{N}\!\!\!\!\!\!\!\int_{\bm{k}}
  \frac{1}{|\bm{k}|^2(E_b-\bm{q}\cdot\hat{\bm{k}})
                     (E_c + \bm{q} \cdot \hat{\bm{k}})} -
  \frac{i}{4(E_b+E_c)}
  \mathcal{N}\!\!\!\!\!\!\!\int_{\bm{k}} \bigg[
\frac{1}{\Delta_c[\bm{k}^2 - (\Delta_c+E_c)^2 -i\varepsilon]}
\nonumber\\
&&
-
   \frac{1}
   {\Delta_b[\bm{k}^2 -(\Delta_b-E_b)^2 -i \varepsilon]} \bigg],
\\
S_3^i
&=&\frac{i}{8} \mathcal{N}\!\!\!\!\!\!\!
\int_{\bm{k}} \frac{k^i}{|\bm{k}|^3(E_b-\bm{q}\cdot
  \hat{\bm{k}})(E_c + \bm{q} \cdot \hat{\bm{k}})}
+ \frac{i}{4(E_b+E_c)} \mathcal{N}\!\!\!\!\!\!\!\int_{\bm{k}}
\bigg[\frac{k^i}{\Delta_c (\Delta_c+E_c)}\;
\nonumber \\
&&
\times
\frac{1}{\bm{k}^2 -(\Delta_c+E_c)^2 -i \varepsilon}
-
\frac{[1+(E_b/\Delta_b)]k^i}
     {[\bm{k}^2 - (\Delta_b-E_b)^2 -i\varepsilon]
      [\bm{k}^2+2\bm{k} \cdot \bm{q} - i \varepsilon]}
\bigg].
\end{eqnarray}
$S_3^0$ and $S_3^i$ include three terms, which correspond
to $S_{3g}^\mu$, $S_{3c}^\mu$, and $S_{3\bar{b}}^\mu$, respectively.

The integrands of the gluon pole contributions  $S_{3g}^0$ and $S_{3g}^i$,
which are the first terms in $S_3^0$ and $S_3^i$, respectively,
have factors $1/(E_Q\pm\bm{q}\cdot\hat{\bm{k}})$ that expand in powers
of $\bm{q}\cdot\hat{\bm{k}}/E_Q$ producing scaleless factors and the factor
$1/|\bm{k}|^2$ does not generate logarithmic divergence. Therefore,
$S_{3g}^\mu=0$. The second terms of $S_3^0$ and $S_3^i$ are the
charm quark pole contributions $S_{3c}^0$ and $S_{3c}^i$. We can follow
the same procedure that was employed in expanding the integrand for $S_{2c}$
in (\ref{s2cexpnad}) to find that $S_{3c}^\mu=0$.
The last terms in $S_3^0$ and $S_3^i$ are the antibottom quark pole
contributions $S_{3\bar{b}}^0$ and $S_{3\bar{b}}^i$, whose structure is
similar to that of $S_{2\bar{b}}$ in (\ref{s2bexpand}). We find that
$S_{3\bar{b}}^0$ contains only scaleless, power-divergent integrals,
which vanish, and the only nonvanishing contribution is $S_{3\bar{b}}^i$:
\begin{equation}
\label{s3b}%
S^{i}_{3 \bar{b}}
=
\frac{i\,{q}^i}{4 (E_b+E_c)\bm{q}^2} \, n_1
=
- \frac{q^i}{16 \pi(E_b+E_c)|\bm{q}|}.
\end{equation}
The Lorentz covariant expression for $S_3^\mu$ is, then, obtained as
\begin{equation}
\label{s3-final}
S_3^\mu
= \frac{i}{(4\pi)^2} \frac{i\pi}{4\delta p^2}
\left[-\frac{p\cdot q}{p^2}p^\mu + q^\mu \right].
\end{equation}
\subsection{$\bm{S_4^{\mu\nu}}$}
The integral $S_4^{\mu\nu}$ is the sum of three contributions:
$S_3^\mu=S_{4g}^{\mu\nu}+S_{4c}^{\mu\nu}+S_{4\bar{b}}^{\mu\nu}$, where
$S_{4g}^{\mu\nu}$, $S_{4c}^{\mu\nu}$, and $S_{4\bar{b}}^{\mu\nu}$ are
the contributions from the poles of the gluon, charm quark, and antibottom
quark, respectively.

After evaluating the $k^0$ integral by contour integration, we find that
\begin{subequations}
\label{s4}
\begin{eqnarray}
S_4^{00} &=&
\frac{i}{8}\mathcal{N}\!\!\!\!\!\!\!\int_{\bm{k}}
\frac{1}{|\bm{k}|(E_b-\bm{q}\cdot \hat{\bm{k}})
                 (E_c + \bm{q} \cdot \hat{\bm{k}})} +
  \frac{i}{4(E_b+E_c)} \mathcal{N}\!\!\!\!\!\!\!\int_{\bm{k}}
   \bigg[
  \frac{1+(E_c/\Delta_c)}
       {\bm{k}^2 - (\Delta_c+E_c)^2 -i\varepsilon}
   \nonumber \\&&
-\frac{1-(E_b/\Delta_b)}
{\bm{k}^2 - (\Delta_b-E_b)^2 -i \varepsilon} \bigg],
\\
S_4^{0i}&=&
- \frac{i}{8}\mathcal{N}\!\!\!\!\!\!\!\int_{\bm{k}}
 \frac{k^i}{|\bm{k}|^2(E_b-\bm{q}\cdot
  \hat{\bm{k}})(E_c + \bm{q} \cdot \hat{\bm{k}})}
-  \frac{i}{4(E_b+E_c)}
  \mathcal{N}\!\!\!\!\!\!\!\int_{\bm{k}}
\bigg[
  \frac{k^i}{\Delta_c[\bm{k}^2 - (\Delta_c+E_c)^2 -i\varepsilon]}
\nonumber\\
&&
-\frac{k^i}
{\Delta_b[\bm{k}^2 -(\Delta_b-E_b)^2 -i \varepsilon]}
\bigg],
\\
S_4^{ij} &=&
\frac{i}{8}\mathcal{N}\!\!\!\!\!\!\!\int_{\bm{k}}
\frac{k^i k^j}{|\bm{k}|^3(E_b-\bm{q}\cdot\hat{\bm{k}})
                         (E_c+\bm{q}\cdot \hat{\bm{k}})} +
  \frac{i}{4(E_b+E_c)} \mathcal{N}\!\!\!\!\!\!\!\int_{\bm{k}}
\bigg[
  \frac{k^i k^j/[\Delta_c (\Delta_c+E_c)]}
       {
        [\bm{k}^2 -(\Delta_c+E_c)^2 -i \varepsilon]}
\nonumber \\&&
 - \frac{[1+(E_b/\Delta_b)]k^ik^j}
   {[\bm{k}^2 - (\Delta_b-E_b)^2 -i\varepsilon]
    [\bm{k}^2+2\bm{k} \cdot \bm{q} - i\varepsilon]}
\bigg].
\end{eqnarray}
\end{subequations}
Like $S_3^\mu$, the three terms in each of $S_4^{00}$, $S_4^{0i}$,
and $S_4^{ij}$ correspond to $S_{4g}^{\mu\nu}$,
$S_{4c}^{\mu\nu}$, and $S_{4\bar{b}}^{\mu\nu}$, respectively.
Following the same procedure to expand the integrands of $S_3^\mu$,
we find that
\begin{eqnarray}
\label{s4b}%
{S}^{ij}_{4\bar{Q}}
&=&
-\frac{i}{2(E_b+E_c)(d-2)}
\left[
\delta^{ij}\left( n_1 -\frac{n_3}{4 \bm{q}^2} \right)
-\frac{{q}^i {q}^j}{\bm{q}^2}\left(n_1-\frac{(d-1)n_3}{4\bm{q}^2}  \right)
\right]
\nonumber \\
&=&
\frac{|\bm{q}|}{32\pi(E_b+E_c)}
\left(
\delta^{ij}
+\frac{{q}^i {q}^j}{\bm{q}^2}
\right).
\end{eqnarray}
The Lorentz covariant expression for $S_{4}^{\mu\nu}$ is now obtained as
\begin{equation}
\label{s4-final}
S_4^{\mu\nu} =\frac{i}{(4\pi)^2} \frac{i\pi \delta}{2}
\bigg\{
  g^{\mu\nu} -
  \bigg[1-\frac{(p \cdot q)^2}{4\delta^2p^4}\bigg] \frac{p^\mu p^\nu}{p^2}
  +\frac{q^\mu q^\nu}{4\delta^2p^2} - \frac{p \cdot q}{4\delta^2p^4}(p^\mu
  q^\nu+ p^\nu q^\mu)
  \bigg\}.\phantom{xxxx}
\end{equation}


\end{document}